\documentclass[11pt]{article} 

\usepackage[utf8]{inputenc} 
\usepackage[english]{babel}
\usepackage{geometry}
\usepackage{amsmath,amsfonts,amsthm,amssymb}
\usepackage{caption}
\usepackage{comment}
\usepackage{setspace}
\usepackage{geometry}
\usepackage{Tabbing}
\usepackage{lastpage}
\usepackage{tabularray}
\usepackage{extramarks}
\usepackage{chngpage}
\usepackage{systeme}
\usepackage{soul}
\usepackage{graphicx,float,wrapfig}
\usepackage[ruled,vlined,linesnumbered]{algorithm2e}
\usepackage{url}
\usepackage{dsfont}
\usepackage{multicol}
\usepackage{tikz}
\usepackage{tablefootnote}
\usetikzlibrary{trees,snakes}
\usetikzlibrary{arrows,automata}
\usetikzlibrary{matrix}
\usetikzlibrary{calc}

\usepackage{longtable} 

\usepackage{makecell} 
\usepackage{tabularx} 
\usepackage{booktabs,threeparttable} 
\usepackage{multirow} 

\usepackage{subcaption} 
\usepackage{mathtools}

\usepackage[ruled]{algorithm2e}
\usepackage{enumerate}
\usepackage{authblk}

\usepackage{natbib}
\usepackage{hyperref}
\usepackage{cleveref} 

\DeclareMathOperator{\vect}{vec}

\title{Dynamic factor analysis for sparse and irregular longitudinal data: an application to metabolite measurements in a COVID-19 study}

\author[1]{Jiachen Cai, Robert J. B. Goudie, Brian D. M. Tom}

\affil[1]{MRC Biostatistics Unit, University of Cambridge, UK}

\date{}

\begin{document}

\maketitle



\begin{abstract}
{It is of scientific interest to identify essential biomarkers in biological processes underlying diseases to facilitate precision medicine. Factor analysis (FA) has long been used to address this goal: by assuming latent biological pathways (statistically, `latent factor') drive the activity of measurable biomarkers (`observed variable'), a biomarker is more influential if its absolute factor loading is larger. Although correlation between biomarkers has been properly handled under this framework, correlation between latent pathways are often overlooked, as one classical assumption in FA is the independence between factors. However, this assumption may not be realistic in the context of pathways, as existing biological knowledge suggests that pathways interact with one another rather than functioning independently. Motivated by sparsely and irregularly collected longitudinal measurements of metabolites in a COVID-19 study of large sample size, we propose a dynamic factor analysis model that can account for the potential cross-correlations between pathways, through a multi-output Gaussian processes (MOGP) prior on the factor trajectories. To mitigate against overfitting caused by sparsity of longitudinal measurements, we introduce a roughness penalty upon MOGP hyperparameters and allow for non-zero mean functions. To estimate these hyperparameters, we develop a stochastic expectation maximization (StEM) algorithm that scales well to the large sample size. In our simulation studies, StEM leads across all sample sizes considered to a more accurate and stable estimate of the MOGP hyperparameters than a comparator algorithm used in previous research, and on average it is $20$ times faster. Application to the motivating example identifies a kynurenine pathway that affects the clinical severity of patients with COVID-19 disease. In particular, a novel biomarker taurine is discovered, which has been receiving increased attention clinically, yet its role was overlooked in a previous analysis of these data. We developed an R package DFA4SIL to implement the proposed method.}


\end{abstract}

\section{Introduction}
\label{background}
A single complex disease or multiple related diseases often represent a dense spectrum of innumerable, overlapping phenotypes resulting from the contribution of a (modest) number of important pathological mechanisms. In the field of precision medicine, the increasing availability of high-dimensional and longitudinal molecular profiling provides opportunities for improving understanding of these mechanisms underlying diseases, which in turn will facilitate more accurate diagnosis and prognosis, and more targeted treatment.

Specifically, this work is motivated by longitudinal measurements of $p = 35$ metabolites (the molecular biomarkers) in a COVID-19 study \citep{ruffieux2023patient}. In this study, blood samples of $n = 101$ SARS-CoV-2 polymerase chain reaction (PCR) positive patients with clinical symptoms were collected at least twice during the 7-week window after symptom onset. These measurements are sparse for each patient, with the subject-specific number of time points ranging between $2$ and $5$, and approximately $50\%$ of patients having only $2$ observations. Moreover, the observation times are irregularly distributed, with $50$ unique time points in total. We aim to identify the key metabolites involved in the biological response to virus infection.

To address this question, Ruffieux et al \citep{ruffieux2023patient} adopted univariate linear mixed models that take each metabolite as the dependent variable separately, and identified a metabolic signature of COVID-19 patients, characterised by increased expression of several intermediates (3-hydroxykynurenine, kynurenine, quinolinic acid) from the kynurenine pathway and decreased expression of the upstream amino acid tryptophan. However, biomarkers often function collaboratively during biological processes underlying complex diseases such as COVID-19. Therefore, instead of characterizing the disease at the individual biomarker level, it may be better to characterize the disease at the level of pathways \citep{richardson2016statistical}, a functional group of biomarkers that work together. To put this pathway perspective into practice, one powerful statistical framework that can be used is factor analysis (FA). Specifically, FA aims to summarize a large number of relevant observed variables (e.g., correlated biomarkers) into a smaller set of latent factors (drivers or pathways) that describe the variability amongst them in a compact manner with limited loss of information. Examples of the use of FA to discover latent biological structure abound in biomedical research, including in chronic diseases such as cancer, dementia, and acute viral infections \citep{chen2011predicting, carvalho2008high, tadde2020dynamic}.

In the case of longitudinal data, dynamic factor analysis approaches~\citep{chen2011predicting, velten2022identifying} have been developed to model time-varying pathway-level responses. However, most approaches assume independence between different factor trajectories; which would translate biologically to assuming that different biological pathways work independently of each other without any interactions. This may not accurately reflect existing biological knowledge, as pathways often work collectively to achieve a complex biological function \citep{hsu2012discovering}, suggesting the need to allow for potential cross-correlations between factor trajectories. Recently, Cai et al \citep{cai2023dynamic} developed a method to relax this assumption. They demonstrated that, by the incorporation of cross-correlations between factors, performance gains were realised in recovering the shape of the underlying pathways' trajectories, uncovering and better interpreting relationships between biomarkers and pathways, and predicting the future trajectories of biomarkers. However, their method is not entirely applicable to our motivating example here. First, the longitudinal biomarkers were measured frequently in the application considered in Cai et al \citep{cai2023dynamic}, with the majority of the individuals having measurements at $16$ regular time points. In contrast, in many biomedical applications, including the motivating example here, measurements/visits are sparse and irregular across subjects. It is unclear how Cai et al's method performs in this setting. Second, Cai et al's application has a small sample size, with $n = 17$; whereas in many applications a larger sample size is anticipated and so an approach that scales to larger sample sizes is necessary.

To bridge the gap, we develop a dynamic factor analysis model that not only permits correlations between factors, but also can handle irregular and sparse longitudinal biomarkers in larger sample size studies. Specifically, the proposed model maps the observed biomarker expression into the latent pathway-level expression via a Bayesian sparse factor analysis (BSFA) and models the potentially correlated pathway trajectories using  multi-output Gaussian processes (MOGP). To mitigate against the issue of over-fitting due to sparsely collected data, we allow for non-zero mean functions on the MOGP and penalize the hyperparameters that control smoothness. For estimating MOGP hyperparameters, we develop a stochastic expectation maximization (StEM) \citep{celeux1985sem} algorithm that we will show scales well. StEM also naturally handles the irregularity of longitudinal measurements across subjects, as it is based on sampling methods. We have developed an R package DFA4SIL (dynamic factor analysis for sparse and irregular longitudinal data) that implements the proposed method.

The remaining paper is structured as follows. Section \ref{model} introduces our proposed model that targets sparse and irregular longitudinal measurements of biomarkers. Section~\ref{inf} discusses inference, which comprises of a StEM algorithm for estimating MOGP hyperparameters and a Gibbs sampler for other variables in the model. In Section~\ref{performance_improvement_sim}, we describe our simulation study that assesses the performance of StEM under different sample sizes, with comparison to the Monte Carlo expectation maximization (MCEM) algorithm used by Cai et al \citep{cai2023dynamic}. Section \ref{application} applies our proposal to the motivating COVID-19 data, and discusses biological implications of results.

\section{Model}
\label{model}

\subsection{Uncovering the sparse latent structure via BSFA}
FA connects observed data to latent underlying factors via factor loadings. Mathematically, let $x_{ijg}$ denote the $g$th biomarker expression of the $i$th subject at the $j$th time point, $y_{ija}$ denote the (latent) $a$th pathway expression of the $i$th subject at the $j$th time point, and $l_{ga}$ represent the loading of the $g$th biomarker on the $a$th pathway. Then the FA model corresponds to 
\begin{equation}
     x_{ijg} = \mu_{ig} + \sum_{a=1}^{k}l_{ga}y_{ija} + e_{ijg},
\label{factor_analysis_scaler_representation}
\end{equation}
where $\mu_{ig}$ is the intercept term for the $g$th biomarker of the $i$th individual (hereafter the ``subject-biomarker mean''), $e_{ijg}$ is the residual error assumed to follow a normal distribution $\text{N}(0, \phi_g^2)$; $i= 1,...,n$; $j = 1,...,q_i$; $g = 1,...,p$; and $a = 1,...,k$, where $n$, $q_i$, $p$, $k$ denote the number of subjects, subject-specific number of observed time points, number of biomarkers and latent factors, respectively. Note that our method allows for subject-specific observed times $\mathbf{t}_{i}=\cup_{j=1}^{q_i}t_{ij}$ for the $i$th subject, which is a key feature of the motivating example that our method aims to accommodate. Throughout this work, the number of factors $k$ is assumed to be pre-specified based on prior knowledge. We recommend, in practice, comparing results across several $k$s to see if different insights would emerge and then choosing the most appropriate $k$.

The factor loading $l_{ga}$ characterizes the relationship between the $g$th biomarker and the $a$th pathway. A larger absolute value suggests the biomarker is more important in relation to the pathway, whereas $l_{ga} =0$ indicates no association at all. We assume only a small proportion of biomarkers are involved in each pathway, an assumption that is often supported by prior biological knowledge. This assumption implies that for any pathway, say the $a$th, most of the $l_{ga}, g = 1,...,p$ are zeroes (i.e., mathematical sparsity). Under the Bayesian framework, we desire a prior that promotes shrinkage of $l_{ga}$ to exactly $0$, so that we do not need to artificially set a threshold for determining whether to include the $g$th biomarker as a member of the $a$th pathway. Out of this consideration, we choose the point-mass mixture prior \citep{carvalho2008high} for $l_{ga}$, which can be expressed as follows. First, we decompose $l_{ga}$ as a product of a binary variable $Z_{ga}$ and a continuous variable $A_{ga}$, then assign a Bernoulli-Beta prior for $Z_{ga}$ and a Normal-Inverse-Gamma prior for $A_{ga}$: 
\begin{equation} 
\label{factor_loading_model}
\begin{split}
    l_{ga} & = Z_{ga} \cdot A_{ga} \\
{Z}_{ga} \sim \text{Bern}(\pi_a), & \;\ \pi_a \sim \text{Beta}(c_0, d_0) \\
{A}_{ga} \sim \text{N}(0,\rho_a^2),  & \;\ \rho_a^2 \sim \text{Inverse-Gamma}(c_1, d_1)
\end{split}
\end{equation}
where $g=1,\ldots,p$, $a=1,\ldots,k$; and $c_0, d_0, c_1, d_1$ are pre-specified positive constants. $Z_{ga}$ indicates the inclusion status of the $g$th biomarker on the $a$th pathway: $Z_{ga} = 1$ means the $g$th biomarker relates to the $a$th pathway, $Z_{ga} = 0$ means it does not. Prior belief about sparsity can be represented via the hyperparameters $c_0$ and $d_0$, which control the proportion of biomarkers $\pi_a$ associated with the $a$th pathway. Finally, we complete the model specification by assigning a Normal-Inverse-Gamma prior to subject-biomarker means $\mu_{ig}$ and variances $\phi_{g}$ in Equation~\ref{factor_analysis_scaler_representation}: 
\begin{align*}
\mu_{ig} & \sim \text{N}(\mu_g, \sigma_g^2),& \sigma_g^2 &\sim\text{Inverse-Gamma}(c_2, d_2); &\ 
\phi_g^2 &\sim\text{Inverse-Gamma}(c_3, d_3),
\end{align*}
where $\mu_g$ is fixed to the mean of the $g$th biomarker expression across all time points of all subjects, and $c_2, d_2, c_3, d_3$ are pre-specified positive constants.

\subsection{Modelling correlated factor trajectories using MOGP}
We model the latent pathway trajectories using Gaussian processes (GP), which have been widely used for analysing functional data \citep{shi2011gaussian}. More specifically, to reflect our prior belief that multiple pathway (factor) trajectories may be correlated, we adopt multi-output Gaussian processes (MOGP). MOGP provide improved prediction relative to assuming independent GP across outputs \citep{bonilla2007multi, boyle2007gaussian, shi2017regression, cheng2020sparse, jones2022alignment} when the true data generating process induces correlation between outputs. A major challenge when constructing MOGP is to appropriately define cross-covariance functions that imply a positive definite covariance matrix. We adopt the commonly used framework of convolution processes (CP) \citep{boyle2005dependent}. We introduce the CP framework in our context in Section~\ref{cp} and propose in Section~\ref{handle_sparse} methods to avoid overfitting with sparse measurements under CP.

\subsubsection{Convolution process for constructing the MOGP prior on factor trajectories}
\label{cp}
\begin{figure}
\centering
\includegraphics[width=0.4\textwidth] 
{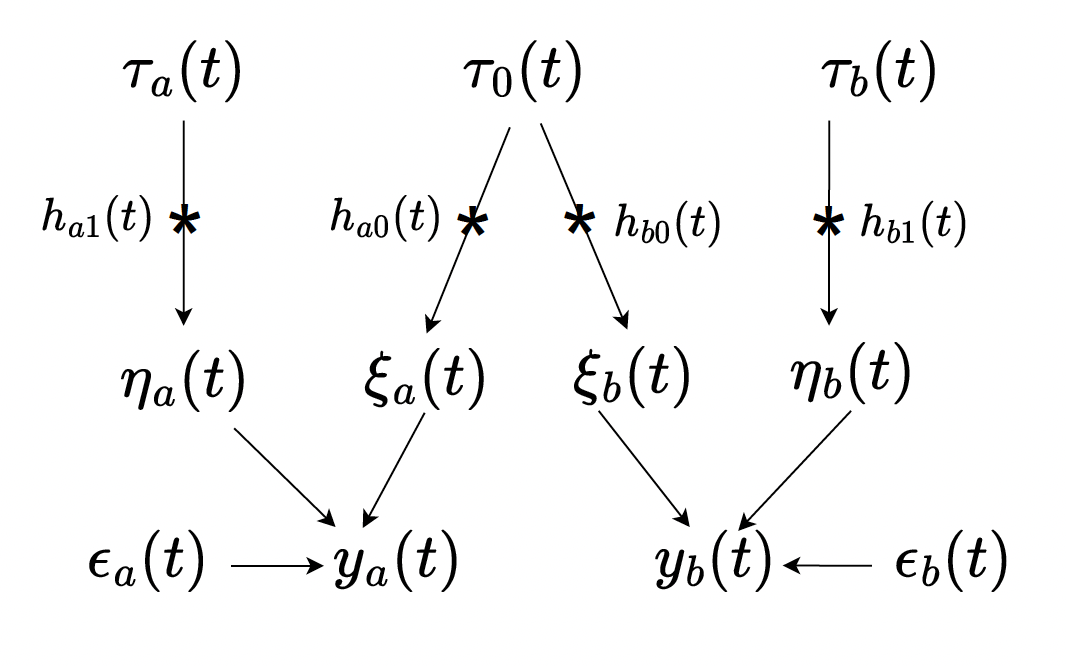}
\caption{Illustration of the kernel convolution framework for MOGP. The star ($\ast$) denotes a convolutional operation and directed arrows indicate direct dependence. $t$ denotes time; $a,b$ are indexes of factor trajectories; $\tau_a(t),\tau_0(t),\tau_b(t)$ are independent Gaussian white noise processes; $h_{a1}(t),h_{a0}(t),h_{b0}(t),h_{b1}(t)$ are Gaussian kernel functions; $\epsilon_a(t),\epsilon_b(t)$ are residuals; $y_{a}(t),y_{b}(t)$ are the $a$th and $b$th factor trajectories, respectively.}
\label{kernel_convolution}
\end{figure}

In the CP framework, correlated processes are generated by introducing common ``base processes''. To introduce the idea, consider the special case of two stochastic processes ${y}_{a}(t)$ and ${y}_{b}(t)$ as an example. In this case, the CP construction, as illustrated in Figure~\ref{kernel_convolution}, is,
\begin{equation}
\begin{split}
\label{factor_score_modeling}
y_{a}(t)& =\eta_a(t)+\xi_a(t)+\epsilon_a(t), \\
y_{b}(t)& =\eta_b(t)+\xi_b(t)+\epsilon_b(t),
\end{split}
\end{equation}
where $\epsilon_a(t), \ \epsilon_b(t)$ are residual errors drawn from a
$\text{N}(0, \psi^2)$ distribution, and $\eta_a(t), \ \eta_b(t), \ \xi_a(t), \ \xi_b(t)$ are processes constructed as follows. 
%
First, three independent, zero-mean base processes $\tau_0(t)$, $\tau_a(t)$ and $\tau_b(t)$ are introduced, which are all Gaussian white noise processes. The first process $\tau_0(t)$ is shared by both ${y}_{a}(t)$ and ${y}_{b}(t)$, thereby inducing dependence between them; whereas $\tau_a(t)$ and $\tau_b(t)$ are specific to ${y}_{a}(t)$ and ${y}_{b}(t)$, respectively, and are responsible for capturing the unique aspects of each process. %
Second, Gaussian kernel functions $h_{a0}(t),\ h_{a1}(t),\ h_{b0}(t),\ h_{b1}(t)$ are applied to convolve the base processes: with $h_{-0}(t)$ applied to the shared process $\tau_0(t)$ and $h_{-1}(t)$ to the output-specific processes $\tau_a(t)$ and $\tau_b(t)$:
\begin{align*}
 \xi_a(t)&=h_{a0}(t)*\tau_{0}(t), & \eta_a(t)&=h_{a1}(t)*\tau_a(t), \\
  \xi_b(t)&=h_{b0}(t)*\tau_{0}(t), & \eta_b(t)&=h_{b1}(t)*\tau_b(t),
\end{align*}
where the convolution operator $*$ is defined as $h(t) * \tau(t) = \int^{\infty}_{-\infty} h(t-s)\tau(s)ds $. All kernel functions $h(t)$ considered here take the form $h(t) = v\,\text{exp}\{-\frac{1}{2}{B}t^2\}$, where $v$ and ${B}$ are parameters that are specific to each kernel function; $B>0$. Note the model in Equation~\ref{factor_score_modeling} can be easily extended to more than two processes. A simple way is as follows \citep{shi2011gaussian},
\begin{equation*}
\begin{split}
y_{a}(t)& =\eta_a(t)+\xi_a(t)+\epsilon_a(t), \ \text{for} \ a=1,...,k, \\
\end{split}
\end{equation*}
where $k$ is the number of processes; $\eta_a(t)$, $\xi_a(t)$, and $\epsilon_a(t)$ are defined in a similar way as above.

In our context, we will associate each pathway trajectory with a component of the MOGP. That is, we assume a MOGP prior for the vector of $k$ pathway trajectories of the $i$th subject, 
\begin{align}
\label{dgp_model}
\begin{split}
        \left(y_{i1}(t),...,y_{ik}(t)\right) \sim \text{MOGP}\left(\mathbf{0}, g(\boldsymbol{\Theta})\right),
\end{split}
\end{align}
where $y_{ia}(t)$ denotes the $a$th factor trajectory of the $i$th subject. The mean function of MOGP is temporarily assumed to be $\mathbf{0}$ following common practice, though we will relax this assumption later in Section~\ref{handle_sparse} as one of the measures to handle the sparse measurements. The kernel function $g(\boldsymbol{\Theta})$ is dependent upon a set of MOGP hyperparameters $\boldsymbol{\Theta} = \{\mathbf{v}, \mathbf{B}\}$, the specific form of which will be discussed in Section~\ref{handle_sparse}. Note that we have assumed a common prior shared by all subjects, as we do not wish to differentiate subjects {\em{a priori}}. Nevertheless, the posterior pathway trajectories (after observing subject-specific biomarker expression) are subject-specific, which we anticipate will reveal the distinction between subjects with different clinical outcomes. 

The prior in Equation~\ref{dgp_model} induces a multivariate normal (MVN) prior distribution for the $k$ latent factors at a finite set of observed time points. Specifically, for the $i$th subject with $q_i$ measurements, the $kq_i$-dimensional vector has the distribution
\begin{align}
\begin{split}
(y_{i11},...,y_{iq_i1},......,y_{i1k},...,y_{iq_ik})^{T} \sim \text{MVN}\left(\mathbf{0}, \Sigma_{\mathbf{Y}_i}(\boldsymbol{\Theta}, \mathbf{t}_i)\right),
\end{split}
\end{align}
where $\mathbf{Y}_i = (\mathbf{y}_{i1},...,\mathbf{y}_{ik})^{T} \in \mathbb{R}^{k\times q_i}$ is the matrix of pathway expression of the $i$th individual, with $\mathbf{y}_{ia}= (y_{i1a}, ..., y_{iq_ia})^{T}$ denoting the $a$th factor's expression across observation times $\mathbf{t}_i$. Below, we will use the notation $\vect({\mathbf{Y}_{i}}^{T})$ to denote the column vector $(y_{i11},...,y_{iq_i1},......,y_{i1k},...,y_{iq_ik})^{T}$ that is a vectorised form of the matrix ${\mathbf{Y}_{i}}^{T}$. The matrix $\Sigma_{\mathbf{Y}_i}(\boldsymbol{\Theta}, \mathbf{t}_i) \in \mathbb{R}^{kq_i \times kq_i}$ is the covariance matrix of $\vect({\mathbf{Y}_{i}}^{T})$, which is dependent on the kernel function $g(\boldsymbol{\Theta})$ and observation times $\mathbf{t}_i$.


\subsubsection{Handling sparse measurements using penalized MOGP with non-zero mean functions}
\label{handle_sparse}
While GP work well with a moderate or large amount of data, in our setting with sparse data, GP (Equation~\ref{dgp_model}) are prone to overfitting \citep{topa2012gaussian, mohammed2017over, manzhos2023rectangularization}, resulting in unreasonably ``wiggly'' fitted curves. To address this issue, regularization is necessary: specifically, we will constrain the hyperparameters that control smoothness and avoid assuming a zero-mean function prior.

\begin{figure}[htp]
\centering
\includegraphics[width=\textwidth]{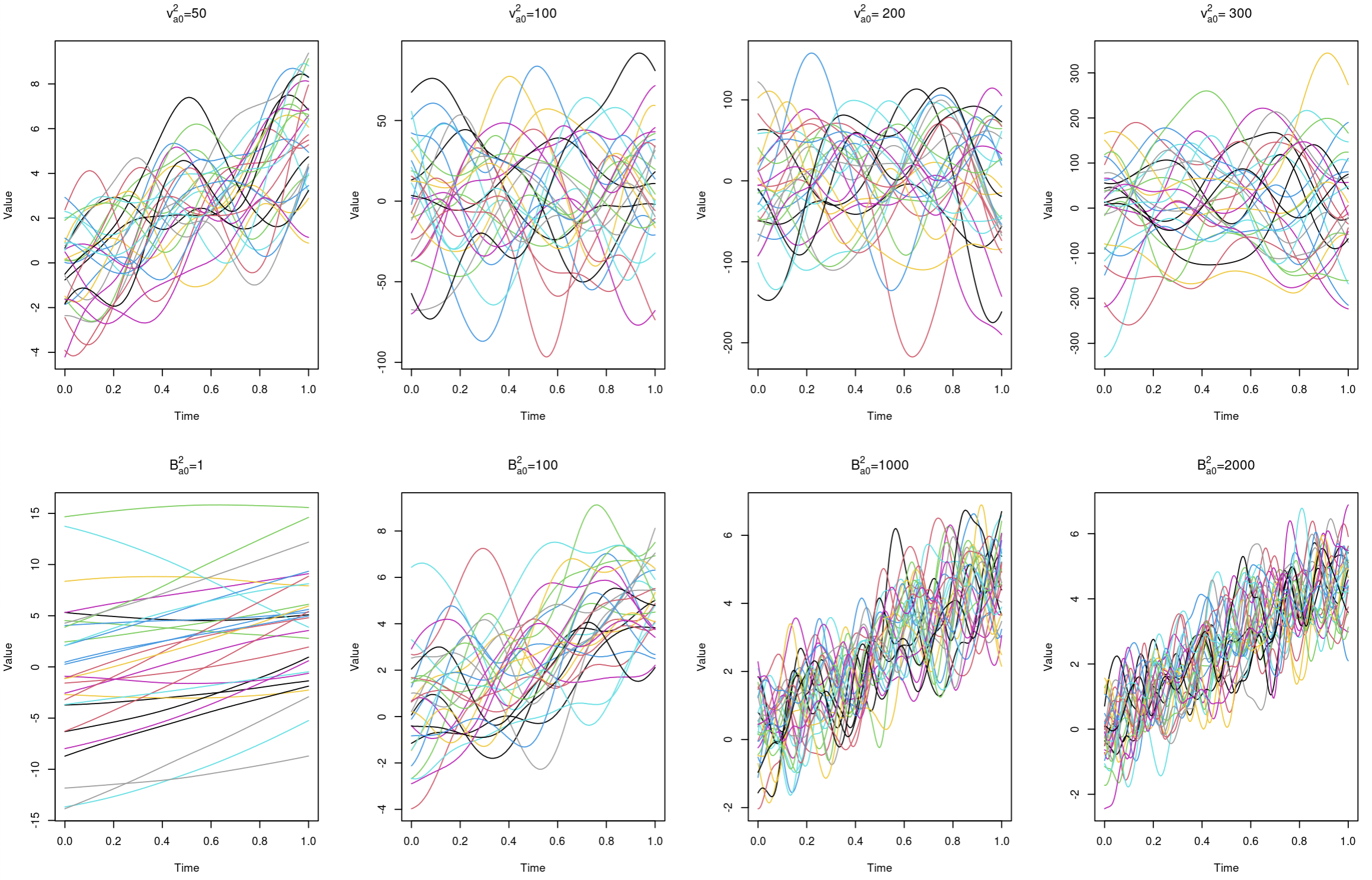}
\caption{$30$ sample curves generated from MOGP under different values of $v_{a0}$ and $B_{a0}$. Note that when varying one hyperparameter, the remaining hyperparameters are fixed.}
\label{dgp_property}
\end{figure}

Under the CP framework introduced in Section~\ref{cp}, the auto covariance function between the time $t_j$ and $t_\ell$ within a single process $a$, denoted as $C_{aa}^{Y}(t_j, t_\ell)$, takes the following form, 
\begin{align}
\label{auto_covariance_expression}
    \begin{split}
         C_{aa}^{Y}(t_j,t_\ell)&=C_{aa}^{\xi}(t_j,t_\ell)+C_{aa}^{\eta}(t_j,t_\ell)+\delta_{j\ell}\psi^2,\\
        C_{aa}^{\xi}(t_j,t_\ell)&=v_{a0}^2\,\frac{\pi^{\frac{1}{2}}}{{B_{a0}}}\,\text{exp}\left\{-\frac{1}{4}B_{a0}^2 d_t^2\right\}, \\
        C_{aa}^{\eta}(t_j,t_\ell)&=v_{a1}^2\,\frac{\pi^{\frac{1}{2}}}{{B_{a1}}}\,\text{exp}\left\{-\frac{1}{4}B_{a1}^2 d_t^2\right\},\\
    \end{split}
\end{align}
where $d_t=t_j-t_\ell$; $\delta_{j\ell}=1$ if $j=\ell$, otherwise $\delta_{j\ell}=0$; $B_{a0}$, $B_{a1}$ are positive parameters. This expression for the auto covariance function indicates that $v_{a0}$ and $v_{a1}$ control the amplitudes of MOGP, while $B_{a0}$ and $B_{a1}$ determine smoothness, as they regulate how quickly the correlation decreases as the absolute time difference $|d_t|$ increases. Figure~\ref{dgp_property} illustrates how the smoothness changes with these hyperparameters. Since $v_{a0}$ and $v_{a1}$ have similar behaviour, and $B_{a0}$ and $B_{a1}$ have similar behaviour, we present results under varying $v_{a0}$ and $B_{a0}$ only. It is clear that larger values of $B_{a0}$ and $B_{a1}$ correspond to rougher curves. This observation motivates the roughness penalty that we will introduce in Section~\ref{inf_mogp}.

In addition to the aforementioned constraints on the covariance function, we also adopt non-zero mean functions for MOGP to handle the sparse measurements. GP with zero mean functions can approximate arbitrary continuous functions, if given enough data, and so zero-means are often used in practice. However, in our setting with few observations, GP with zero mean functions predict zeroes in regions with little data, which is problematic \citep{iwata2022few}. The importance of the prior mean function when data are sparse has been highlighted in previous research as potentially affecting both the prediction performance \citep{fortuin2019meta} and the estimation of hyperparameters of the covariance function  \citep{hwang2023use}. Specifically, we assume a constant mean function $c_a$ for the $a$th pathway, $a = 1,...,k$, and will estimate $c_a$ from the data, along with the hyperparameters $\boldsymbol{\Theta}$ that determine the covariance function.


\subsection{Proposed model}

Finally, we present the proposed model in its matrix form. Let $\mathbf{X}_i=(\mathbf{x}_{i1},...,\mathbf{x}_{ip})^{T} \in \mathbb{R}^{p \times q_i}$ be the matrix of biomarker expressions at $q_i$ observation times for the $i$th individual, with $\mathbf{x}_{ig}= (x_{i1g}, ..., x_{iq_{i}g})^{T}$ denoting the $g$th biomarker's longitudinal measurements. Correspondingly, let $\mathbf{M}_{i} = (\boldsymbol{\mu}_{i1}, ..., \boldsymbol{\mu}_{ip})^{T} \in \mathbb{R}^{p \times q_i}$ be the matrix of subject-biomarker means, with $\boldsymbol{\mu}_{ig}=\mu_{ig}\mathbf{1}_i$, where $\mathbf{1}_i$ is a $q_i$-dimensional column vector consisting of the scalar $1$; $\mathbf{C}_i = (\mathbf{c}_1^T,...,\mathbf{c}_k^T)^{T} \in \mathbb{R}^{kq_i}$ is the mean vector for $\vect({\mathbf{Y}_{i}}^{T})$, with $\mathbf{c}_a=c_{a}\mathbf{1}_i$. Furthermore, let $\mathbf{L}=\{l_{ga}\}_{g = 1,..., p, a = 1,...,k} \in \mathbb{R}^{p \times k}$ be the matrix of factor loadings, $\mathbf{A}=\{A_{ga}\}_{g=1,...,p; a = 1,...,k} \in \mathbb{R}^{p\times k}$ be the matrix of regression coefficients and $\mathbf{Z}=\{Z_{ga}\}_{g=1,...,p; a = 1,...,k} \in \mathbb{R}^{p\times k}$ be the matrix of inclusion indicators,
\begin{equation}
\label{integrated_model}
\begin{split}
        \mathbf{X}_i&=\mathbf{M}_{i} + \mathbf{L}\mathbf{Y}_{i}+\mathbf{E}_{i},\\
        \mathbf{L}&=\mathbf{A}\circ \mathbf{Z},\\
        \vect({\mathbf{Y}_{i}}^{T}) & \sim \text{MVN}\left(\mathbf{C}_i, \Sigma_{\mathbf{Y}_i}(\boldsymbol{\Theta}, \mathbf{t}_i)\right) \\
\end{split}
\end{equation}
where $\mathbf{E}_i$ is the residual matrix, and $\circ$ denotes element-wise matrix multiplication.

\section{Inference}
\label{inf}
To achieve a balance between the computational cost and statistical performance, we use an empirical Bayes (EB) approach to obtain point estimates of MOGP hyperparameters $\boldsymbol{\Theta}$ and $\mathbf{C}$, where $\mathbf{C}=\{c_a\}_{a=1,...,k}$. Then we implement a Gibbs sampler for other parameters under the fixed estimate of $\boldsymbol{\Theta}$ and $\mathbf{C}$ in order to quantify the uncertainty of estimating quantities of major interest, such as latent factor loadings $\mathbf{L}$ and factor trajectories $\mathbf{Y}_{i}$. 

Specifically, we develop a StEM algorithm \citep{celeux1985sem} in Section~\ref{inf_mogp} to calculate the maximum likelihood estimate (MLE) of $\boldsymbol{\Theta}$ and $\mathbf{C}$ (denoted as $\widehat{\boldsymbol{\Theta}}^{\text{MLE}}$ and $\widehat{\mathbf{C}}^{\text{MLE}}$, respectively) which maximize a penalized likelihood that penalizes MOGP roughness. In Section~\ref{gibbs_sampler}, we describe the Gibbs sampler for all parameters in the model except for $\boldsymbol{\Theta}$ and $\mathbf{C}$, represented as $\boldsymbol{\Omega}=\{\mathbf{M}, \mathbf{Y}, \mathbf{A}, \mathbf{Z}, \boldsymbol{\rho}, \boldsymbol{\pi}, \boldsymbol{\sigma}, \boldsymbol{\phi}\}$, where $\mathbf{M}=\{\mathbf{M}_i\}_{i=1,...,n}$, $\mathbf{Y}=\{\mathbf{Y}_i\}_{i=1,...,n}$, $\boldsymbol{\rho}=\{\rho_a^2\}_{a=1,...,k}$, $\boldsymbol{\pi}=\{\pi_a\}_{a=1,...,k}$,   $\boldsymbol{\sigma}=\{\sigma_g^2\}_{g=1,...,p}$, $\boldsymbol{\phi}=\{\phi_g^2\}_{g=1,...,p}$. This sampler serves two purposes. First, within the StEM algorithm, it simulates samples of $\boldsymbol{\Omega}$ in the stochastic step (S-step) for updating the estimates of $\boldsymbol{\Theta}$ and $\mathbf{C}$ in the maximization step (M-step). Second, after the StEM algorithm, it generates samples of $\boldsymbol{\Omega}$ for posterior inference.

\subsection{Estimating MOGP hyperparameters by maximizing the penalized likelihood}
\label{inf_mogp}
To calculate $\widehat{\boldsymbol{\Theta}}^{\text{MLE}}$ and $\widehat{\mathbf{C}}^{\text{MLE}}$, we maximize the following objective function, 
\begin{align}
\label{penalized_q_specific}
& \ln f(\mathbf{X}\mid\boldsymbol{\Theta}, \mathbf{C}) - \lambda \sum_{a = 1}^{k} ({B_{a0}} + {B_{a1}}),
\end{align}
where $f(\mathbf{X}\mid\boldsymbol{\Theta}, \mathbf{C})$ is the likelihood of observing all biomarker expressions $\mathbf{X} = \{\mathbf{X}_i\}_{i=1,..,n}$, $\lambda$ is the parameter that tunes the extent to which roughness is penalised. The likelihood $f(\mathbf{X}\mid\boldsymbol{\Theta}, \mathbf{C})$ involves high-dimensional integration taking the following form, 
\begin{align}
    \label{eqn:marglike}
         \begin{split} 
f(\mathbf{X}\mid\boldsymbol{\Theta}, \mathbf{C})&=\int{f(\mathbf{X},\boldsymbol{\Omega}\mid\boldsymbol{\Theta}, \mathbf{C})}d\boldsymbol{\Omega}\\
            &=\int{f(\mathbf{X}\mid\mathbf{M}, \mathbf{Y}, \mathbf{A}, \mathbf{Z}, \boldsymbol{\phi})\, f(\mathbf{M}\mid\boldsymbol{\sigma})\, f(\mathbf{Y}\mid\boldsymbol{\Theta}, \mathbf{C}})\,f(\mathbf{A}\mid\boldsymbol{\rho})\,f(\mathbf{Z}\mid\boldsymbol{\pi})\,f(\boldsymbol{\phi})\,f(\boldsymbol{\sigma})\,f(\boldsymbol{\rho})\,f(\boldsymbol{\pi})d\boldsymbol{\Omega},\\
            \end{split}
\end{align}
which cannot be evaluated analytically. 

\subsubsection{Stochastic expectation maximization (StEM)}
\label{stem}
To deal with the integration in Equation~\ref{eqn:marglike}, we consider expectation maximization (EM)-type algorithms; by viewing $\boldsymbol{\Omega}$ as hidden variables and then iteratively constructing a sequence of $\widehat{\boldsymbol{\Theta}}^{(l)}$ and $\widehat{\mathbf{C}}^{(l)}$ that converges to the truth, $l=1,2,3,...$. As a closed-form of the expectation of the complete data log-likelihood is unavailable due to the complexity of the model, we resort to a variant of EM based on simulations \citep{ruth2024review}. The StEM algorithm is particularly suitable for our use here, as it only requires a single random simulation/sample at each iteration \citep{diebolt1995stochastic} and therefore is scalable to the large sample size of the motivating example. StEM has also been found to be less sensitive to initial values than alternative algorithms \citep{svensson2010asymptotic, davies2021stochastic}. Specifically, the risk of getting stuck in local optima is reduced.

The sequence of $\widehat{\boldsymbol{\Theta}}^{(l)}$ generated from StEM constitutes a homogeneous Markov chain (the same applies to the sequence of $\widehat{\mathbf{C}}^{(l)}$), which means that its convergence is with regard to the average of the last $m$ iterated values of the parameter \citep{nielsen2000stochastic}. Therefore, it can be treated like Markov chain Monte Carlo (MCMC) samples \citep{zhang2020improved}. For instance, tuning the burn-in size is essential, and techniques related to assessing MCMC convergence can also be used. Specifically, StEM comprises two steps. In the S-step of the $l$th iteration, one random sample of $\boldsymbol{\Omega}$ (denoted as $\boldsymbol{\Omega}^{(l-1)}$) is generated from the conditional distribution $f(\boldsymbol{\Omega}\mid\mathbf{X}, \widehat{\boldsymbol{\Theta}}^{(l-1)}, \widehat{\mathbf{C}}^{(l-1)})$ to constitute the complete data $(\mathbf{X}, \boldsymbol{\Omega}^{(l-1)})$. Then in the M-step, the update $\widehat{\boldsymbol{\Theta}}^{(l)}$ and $\widehat{\mathbf{C}}^{(l)}$ can be found by maximizing the penalized log-likelihood of observing the complete data, 
\begin{align}
\label{m_step_obj}
& \ln f(\mathbf{X}, \boldsymbol{\Omega}^{(l-1)}\mid\boldsymbol{\Theta}, \mathbf{C}) - \lambda \cdot \sum_{a = 1}^{k} ({B_{a0}} + {B_{a1}})
\end{align}
Applying the factorisation in Equation~\ref{eqn:marglike} to Equation~\ref{m_step_obj} reveals that the only term in $\ln f(\mathbf{X}, \boldsymbol{\Omega}^{(l-1)}\mid\boldsymbol{\Theta}, \mathbf{C})$ that depends on $\boldsymbol{\Theta}$ is $f(\mathbf{Y}^{(l-1)}\mid\boldsymbol{\Theta}, \mathbf{C})$, which immediately reduces the objective function to be maximized in Equation~\ref{m_step_obj} to 
\begin{align*}
\begin{split}
& \ln f(\mathbf{Y}^{(l-1)}\mid\boldsymbol{\Theta}, \mathbf{C}) - \lambda 
\sum_{a = 1}^{k} ({B_{a0}} + {B_{a1}}), \\
\end{split}
\end{align*}
where $f(\mathbf{Y}^{(l-1)}\mid\boldsymbol{\Theta}, \mathbf{C})$ is the product of MVN distributions as a result of the MOGP prior on factor trajectories of all subjects (Equation~\ref{integrated_model}). Expanding this term leads to
\begin{align}
\label{m_step_obj_sim_full}
\begin{split}
& \sum_{i=1}^{n} \ln \text{MVN} \left( \vect({\mathbf{Y}_{i}}^{T})^{(l-1)} \ \middle|\ \Sigma_{\mathbf{Y}_i}(\boldsymbol{\Theta}, \mathbf{t}_i), \mathbf{C_i} \right)  - \lambda \sum_{a = 1}^{k} ({B_{a0}} + {B_{a1}}), \\
\end{split}
\end{align}
which is then maximised in the M-step.

\subsubsection{Adaptations to facilitate StEM implementation via an existing R package}
\label{stem_gpfda}
To find the estimate of $\boldsymbol{\Theta}$ and $\mathbf{C}$ that can maximize Equation~\ref{m_step_obj_sim_full}, we use the  R package `GPFDA' \citep{konzen2021gaussian} (short for Gaussian processes for functional data analysis), which return the MLEs of hyperparameters of the MOGP modelled via the CP framework. GPFDA uses the efficient optimization algorithm L-BFGS, an iterative method based on second-order derivatives. When calculating the MLEs for the longitudinal $\vect({\mathbf{Y}_{i}}^{T})^{(l-1)}$ of a single individual, the computational complexity of GPFDA is $O(k^2q_i^2)$. This complexity is drastically reduced compared to the naive approach of directly inverting the covariance matrix $ \Sigma_{\mathbf{Y}_i}(\boldsymbol{\Theta}, \mathbf{t}_i)$ via a Cholesky decomposition, where the computational complexity would be $O(k^3q_i^3)$. In addition, GPFDA allows for the specification of different types of mean functions, such as the output-specific mean functions that we assume in our model.

However, there are two major limitations that prevent us from directly using the GPFDA package. First, the objective function used by GPFDA is the observed log-likelihood function and it does not allow specification of a penalty term. Second, GPFDA assumes all input processes are measured at common time points (rather than subject-specific observed times $\mathbf{t}_i$), which is neither the case in our motivating example nor in many other practical settings.

To overcome these limitations, we first modified the GPFDA package so that users can specify the penalized objective function in Equation~\ref{m_step_obj_sim_full}. Second, we introduce auxiliary variables to augment the parameter space to enable the use of GPFDA in the setting of irregular longitudinal data. Let $\mathbf{t} = \cup_{i=1}^{n}\mathbf{t}_{i}$ denote the vector of all unique observation times across all subjects and denote its length as $q$, and let $\mathbf{Y}_{i,\text{aug}}  = (\mathbf{y}_{i1},...,\mathbf{y}_{ik})^{T} \in \mathbb{R}^{k \times q}$ denote an augmented version of the matrix $\mathbf{Y}_{i} \in \mathbb{R}^{k \times q_i}$, where $\mathbf{y}_{ia}= (y_{i1a}, ..., y_{iqa})^{T}$ denotes the $a$th factor's expression across all observation times $\mathbf{t}$. 
In this augmented space, the unknown parameters are $\boldsymbol{\Omega}_{\text{aug}} = \{\mathbf{M},\mathbf{Y}_{\text{aug}}, \mathbf{A}, \mathbf{Z}, \boldsymbol{\rho}, \boldsymbol{\pi}, \boldsymbol{\sigma}, \boldsymbol{\phi}\}$, where $\mathbf{Y}_{\text{aug}}=\{\mathbf{Y}_{i,\text{aug}}\}_{i=1,...,n}$. By replacing $\boldsymbol{\Omega}$ with $\boldsymbol{\Omega}_{\text{aug}}$ in Equations~\ref{penalized_q_specific}-\ref{m_step_obj}, the objective function in Equation~\ref{m_step_obj_sim_full} changes accordingly to, 
\begin{align}
\label{m_step_obj_sim_aug}
\begin{split}
\sum_{i=1}^{n} \ln \text{MVN} \left(\vect({\mathbf{Y}_{i, \text{aug}}}^{T})^{(l-1)} \ \middle|\ \mathbf{C}_{\text{aug}}, \Sigma_{\mathbf{Y}_{\text{aug}}}(\boldsymbol{\Theta}, \mathbf{t})\right)  - \lambda \sum_{a = 1}^{k} ({B_{a0}} + {B_{a1}}),\\
\end{split}
\end{align}
where $\mathbf{Y}^{(l-1)}_{i, \text{aug}}  \in \mathbb{R}^{kq}$ denotes a sample of the augmented matrix $\mathbf{Y}_{i,\text{aug}}$, generated by sampling from the appropriate marginal of $f(\boldsymbol{\Omega}_{\text{aug}}\mid\mathbf{X}, \widehat{\boldsymbol{\Theta}}^{(l-1)})$. Such samples can be conveniently obtained using the Gibbs sampler that we will introduce in Section~\ref{gibbs_sampler}. The quantities $\mathbf{C}_{\text{aug}}$ and $\Sigma_{\mathbf{Y}_{\text{aug}}}(\boldsymbol{\Theta}, \mathbf{t})$ are, respectively, the corresponding mean vector and the covariance matrix that are common to all subjects. The modified objective function in Equation~\ref{m_step_obj_sim_aug} after data augmentation enables the estimation of the MLE using GPFDA despite the package permitting only processes measured at common time points. Finally, we point out that the computational complexity of finding the maximizer for Equation~\ref{m_step_obj_sim_aug} is $O(nk^2 q^2)$, constant across all iterations.

\subsubsection{Choice of the tuning parameter $\lambda$}
To choose the value for the tuning parameter $\lambda$ in the penalized log-likelihood function, we use a cross-validation approach to find the value with the smallest averaged mean absolute error (MAE) in predicting biomarker expressions. Specifically, all subjects are first divided randomly into $l$ approximately equal groups. For each candidate value of $\lambda$, the following steps are iterated over each of the $l$ groups,

\begin{enumerate}

\item Use the group corresponding to the current iteration as the test data, and the remaining $(l-1)$ groups as the training data;

\item Fit the model with the penalized log-likelihood in Equation~\ref{penalized_q_specific} to the training data to calculate $\widehat{\boldsymbol{\Theta}}^{\text{MLE}}$;

\item Under $\widehat{\boldsymbol{\Theta}}^{\text{MLE}}$ predict for biomarker expressions in the test data, and compare predictions with the ground truth to calculate MAE;

\end{enumerate}
Finally, the average MAE across all $l$ groups is calculated for the candidate value under consideration.

\subsection{A Gibbs sampler for other variables in the model}
\label{gibbs_sampler}
Under a fixed estimate of MOGP hyperparameters ($\widehat{\boldsymbol{\Theta}}$ and $\widehat{\mathbf{C}}$ denoting estimates of hyperparameters in the covariance function and mean function, respectively), the conditional distribution $f(\boldsymbol{\Omega}_{\text{aug}}\mid \mathbf{X}, \widehat{\boldsymbol{\Theta}}, \widehat{\mathbf{C}})$ does not have a closed-form. To acquire samples from this distribution, we developed an efficient Gibbs sampler, which simulates a Markov chain whose stationary distribution is $f(\boldsymbol{\Omega}_{\text{aug}}\mid\mathbf{X}, \widehat{\boldsymbol{\Theta}}, \widehat{\mathbf{C}})$. Note that to obtain one sample from $f(\boldsymbol{\Omega}_{\text{aug}}\mid\mathbf{X}, \widehat{\boldsymbol{\Theta}}^{(l-1)}, \widehat{\mathbf{C}}^{(l-1)}))$ required by the M-step of StEM for maximization, we run a long chain (the number of iterations $S$ needs to be pre-specified), discard samples in the burn-in period, then take one random sample from the remaining iterations and input it to GPFDA for acquiring $\widehat{\boldsymbol{\Theta}}^{(l)}$ and $\widehat{\mathbf{C}}^{(l)}$. It is worthwhile to run a long chain to generate the single sample required in the M-step because the Gibbs sampler run in the S-step is not time-consuming compared to the M-step where $\widehat{\boldsymbol{\Theta}}^{(l)}$ and $\widehat{\mathbf{C}}^{(l)}$ are calculated. Therefore, running a long chain does not notably affect the computation time.

Full conditionals for all variables in $\boldsymbol{\Omega}_{\text{aug}}$ are available analytically, and the details are available in Supplementary Section A.1. Here, we highlight the updating of the augmented variable $\mathbf{Y}_{\text{aug}}$. When updating $\vect({\mathbf{Y}_{i,\text{aug}}}^{T})$, the vectorized form of the $i$th individual's factor scores at common times $\mathbf{t}$, we adopt a block Gibbs sampler. Specifically, we factorise $f(\vect({\mathbf{Y}_{i,\text{aug}}}^{T})\mid\mathbf{X},\widehat{\boldsymbol{\Theta}}, \widehat{\mathbf{C}}, \boldsymbol{\Omega}_{\text{aug}} \setminus \vect({\mathbf{Y}_{i,\text{aug}}}^{T}))$, according to the partition  $\vect({\mathbf{Y}_{i,\text{aug}}}^{T}) = (\vect({\mathbf{Y}_{i}^{T}}), \vect({\mathbf{Y}_{i,\text{add}}^{T}}))$, where $\boldsymbol{\Omega}_{\text{aug}} \setminus \vect({\mathbf{Y}_{i,\text{aug}}}^{T})$ denotes the remaining parameters excluding $\vect({\mathbf{Y}_{i,\text{aug}}}^{T})$, and $\vect({\mathbf{Y}_{i,\text{add}}^{T}})$ represents the newly added variables. The first factor $f(\vect({\mathbf{Y}_{i}^{T}})\mid\mathbf{X},\widehat{\boldsymbol{\Theta}}, \widehat{\mathbf{C}}, \boldsymbol{\Omega}_{\text{aug}} \setminus \vect({\mathbf{Y}_{i, \text{aug}}}^{T}))$ follows a MVN distribution dependent on observed biomarker expression, as can be seen from the model in Equation~\ref{integrated_model}.
The second factor $f(\vect({\mathbf{Y}_{i,\text{add}}^{T}})\mid\vect({\mathbf{Y}_{i}^{T}}), \widehat{\boldsymbol{\Theta}}, \widehat{\mathbf{C}})$ also follows a MVN distribution according to standard properties of the MOGP model. Our sampler samples from these two factors in turn.

\subsection{Handling non-identifiability}
\label{identifiability}
A common challenge related to factor analysis is non-identifiability. In the specific model we propose here, there are two types of non-indentifiability. First, the covariance matrix for latent factors is not identifiable (see Supplementary Section A.2 for a detailed discussion). To address this non-identifiability, we constrain the main diagonal elements of the covariance matrix to be $1$ (i.e. the variance of each factor at each time point is set to $1$). Second, under the above constraint, factor loadings $\mathbf{L}$ and factor scores $\vect({\mathbf{Y}_{i}}^{T})$ are still not identifiable due to potential sign changes or permutations (we will refer to them together as signed-permutation hereafter). `Sign change' means that the likelihood remains the same if the sign of a factor changes (i.e., from positive to negative, and vice versa). `Permutation' (also known as `label switching' in literature) means that if we shuffle the indices of the factors, the likelihood will not change. To address the signed-permutation associated with $\mathbf{L}$ and $\vect({\mathbf{Y}_{i}}^{T})$, we post-align samples using the R package ``factor.switch'' \citep{papastamoulis2022identifiability}. 

\section{Simulation study}
\label{performance_improvement_sim}
To investigate the performance of the proposed StEM algorithm under different sample sizes, we conducted a simulation study. The data generation mechanism is described in Section~\ref{sim_data_gen}. We compare the results under StEM with those under an alternative algorithm, Monte Carlo expectation algorithm (MCEM), which was used by Cai et al \citep{cai2023dynamic}. Implementation details and results are described in Section~\ref{analysis_methods}.

\subsection{Data generation mechanism}
\label{sim_data_gen}
We followed one data generation mechanism considered in Cai et al \citep{cai2023dynamic}, which mimics longitudinal gene expression data obtained from a human challenge study. We varied the sample size, taking values ranging from $n=10$ to $n=250$. For each sample size, we simulated $100$ datasets according to the setting described below. 

The number of observed time points was fixed at $q_i = 8$ for all individuals, the number of true latent factors was $k = 4$ and the number of biomarkers was $p = 100$. The mean value for each factor score $y_{ija}$ was fixed to $0$, and the covariance matrix $\Sigma_{\mathbf{Y}}$ encodes non-zero cross-correlations between factors (i.e., factors are truly correlated). The exact form of the cross-correlation matrix can be found in Supplementary Figure 8. Each factor was expected to regulate $10\%$ of all biomarkers: we set hyperparameters $c_0 = 0.1 \cdot p$ and $d_0 = 0.9 \cdot p$, leading to $\mathbb{E}\left[\pi_a\right] = 0.1$ (according to Equation~\ref{factor_loading_model}). If a biomarker was regulated by an underlying factor, then the corresponding factor loading was generated from a normal distribution $\text{N}(4,1^2)$; otherwise the factor loading was set to $0$. Each subject-biomarker mean $\mu_{ig}$ was generated from $\text{N}(\mu_g, \sigma^2_g)$, where $\mu_g$ ranged between $4$ and $16$ and $\sigma_g = 0.5$. Finally, observed biomarkers were generated according to Equation \ref{factor_analysis_scaler_representation}, where $\phi_g = 0.5$.

\subsection{Implementation details and results}
\label{analysis_methods}
When implementing the StEM algorithm, we ran $200$ iterations (we monitored the sequence of iterated estimates via traceplots and there were no apparent non-convergence concerns). For the existing MCEM algorithm to which we compare our StEM algorithm, we used the R package DGP4LCF \citep{dgp4lcf}. MCEM stops once the number of increasing the size of Markov chain Monte Carlo samples (as implemented in MCEM by Cai et al \citep{cai2023dynamic}) has reached the pre-defined limit $w$, and the default value in the package is $w=5$; to fully explore its performance, we perform separate runs of MCEM with $w$ set to $5, 8$ and $10$. To assess the statistical accuracy in estimating MOGP hyperparameters $\boldsymbol{\Theta}$, we calculate the mean absolute difference (MAD) between estimated cross-correlations constructed by $\widehat{\boldsymbol{\Theta}}^{\text{MLE}}$ and the ground truth. Note that the (concurrent or lag $0$) cross-correlations between factors
are constant across time, under the CP framework introduced in  Section~\ref{cp}. Specifically, the cross-correlation coefficient between factor $a$ and factor $b$ (denoted as $\rho_{ab}$) is $\rho_{ab} = \frac{C_{ab}^{Y}}{\sqrt{C_{aa}^{Y}C_{bb}^{Y}}}$, where $C_{aa}^{Y}$ and $C_{bb}^{Y}$ are variances of factor $a$ and $b$, respectively. Specific forms can be found in Equation~\ref{auto_covariance_expression}. The covariance between factors $a$ and $b$ is also constant across time, taking the form of $C_{ab}^{Y} =v_{a0}v_{b0}\frac{(2\pi)^{\frac{1}{2}}}{\sqrt{B_{a0}+B_{b0}}}$. 

Under $\widehat{\boldsymbol{\Theta}}^{\text{MLE}}$, we ran the Gibbs sampler three times with different starting values. We ran each chain for $100,000$ iterations, with a $20\%$ burn-in proportion and retained only every $200$th iteration. Due to the identifiability issue caused by sign-permutation (discussed in Section~\ref{identifiability}), we post-aligned samples within and across chains using R package ``factor.switch'' \citep{papastamoulis2022identifiability}. To enable the calculation of MAD, the final estimate also needs to be aligned with the ground truth. To do this, we developed an automatic algorithm to determine the order and sign of final estimated factors. Specifically, the total number of possible signed permutations is $2^k k!$ given $k$ factors; we evaluated the difference between the final estimate and the truth under each possible combination, and ultimately chose the combination that had the smallest MAD (in order to maximize the similarity of all estimates to the ground truth). In addition, we recorded the computational time for each run. We limited computation to a maximum of 36 hours for each simulation. 

\begin{figure}[htp]
\centering
\includegraphics[width = \textwidth]{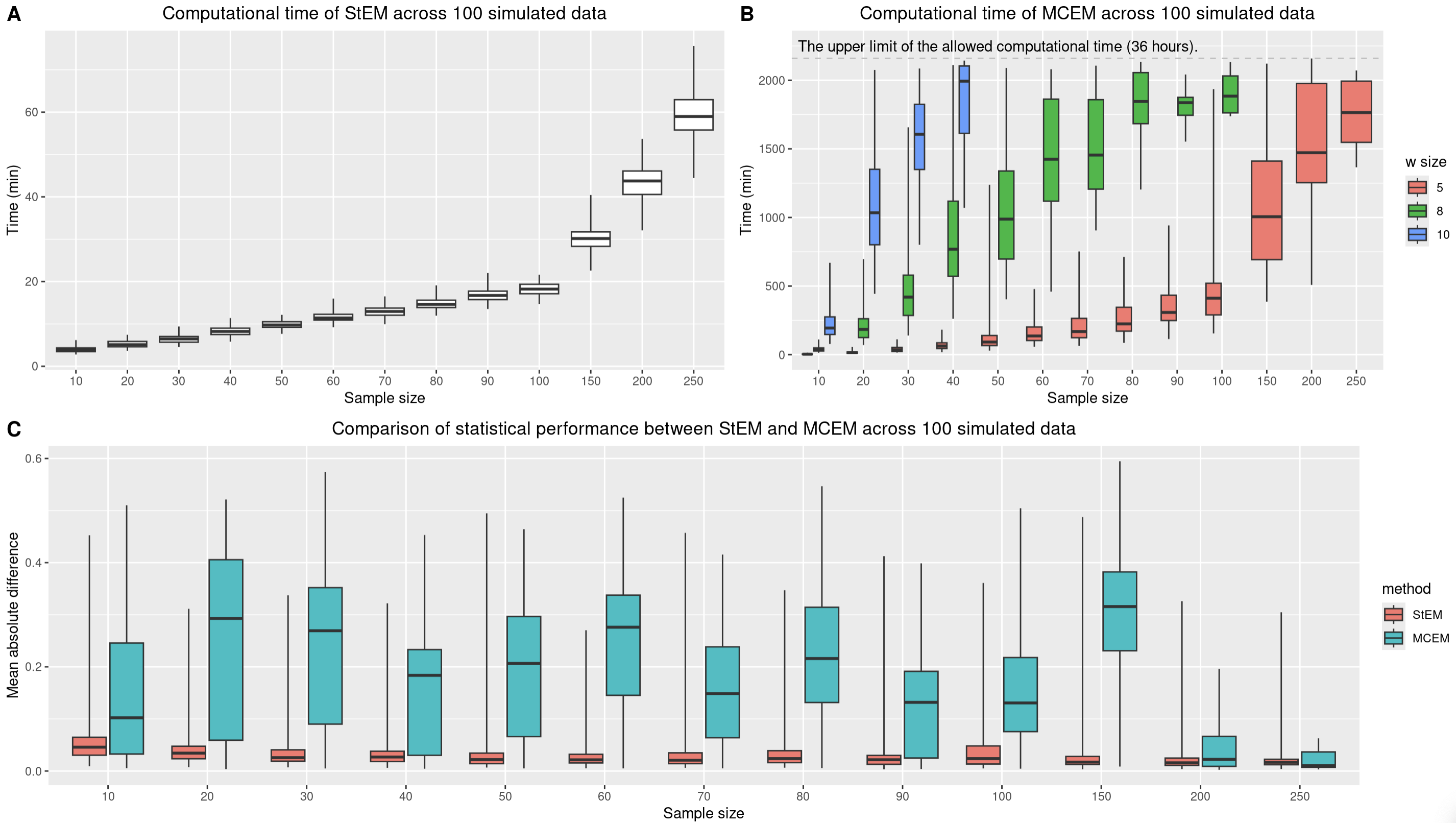}
\caption{Comparison between StEM and MCEM in terms of computational time (Panels A and B) and estimation accuracy (Panel C) under different sample sizes in the simulation study. For each sample size, $100$ dataset were simulated. Whiskers of boxplots extend to the smallest and largest values.}
\label{sim_results}
\end{figure}

Figure~\ref{sim_results} displays results. StEM is much faster than MCEM, regardless of the value of $w$, under all sample sizes (Panels A and B). On average (across sample sizes), StEM is $20$ times faster than MCEM, and this computational gain is more obvious as the sample size increases. With $w=10$, MCEM never returned results within $36$ hours when $n \geq 50$; with $w=8$, MCEM never returned results within $36$ hours when $n \geq 150$. Therefore, corresponding results were not displayed, and when comparing the statistical performance of MCEM with StEM, we focus on the setting $w=5$ for MCEM (Panel C). Overall, estimation performance of StEM generally improves as the sample size $n$ increases, and the median MAD is always better than that of MCEM regardless of $n$. Across all sample sizes, the median MAD under StEM is at least $30\%$ lower than MCEM. In addition, the variability of MAD (across $100$ simulated data) using MCEM is often higher compared to StEM (i.e., StEM has more stable results), which may be due to StEM being less likely to get stuck in local optima, as discussed in Section~\ref{stem}. These results confirm the added value of the StEM development when applying the proposed dynamic factor analysis to studies with larger sample sizes, such as the motivating COVID-19 example.

\section{Data application}
\label{application}

\subsection{Study description}
\begin{figure}[htp]
\centering
\includegraphics[width=\textwidth]{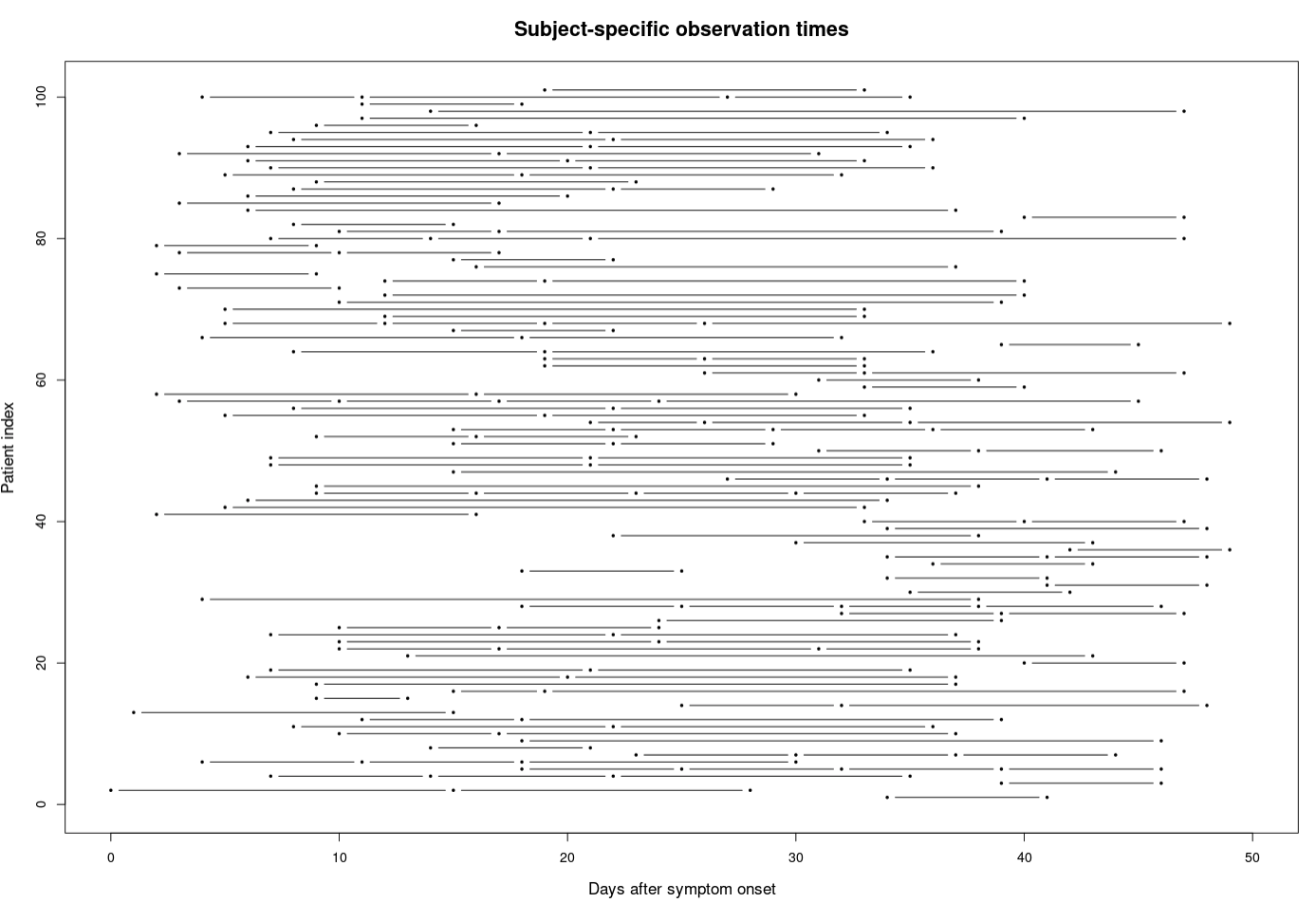}
\caption{Subject-specific observation times. Each line corresponds to a single subject, and connects subject-specific observed times.}
\label{observed_times}
\end{figure}

SARS-CoV-2 PCR-positive patients were recruited at Cambridge University Hospitals between 31st March 2020 and 7th August 2020, and blood samples were collected at and following enrollment to quantify biomarker expressions. Day $0$ is defined as the date of symptom onset, and our analysis focused on the 7-week window after symptom onset. Figure~\ref{observed_times} visualizes subject-specific observed times when the metabolite samples were collected during this period. The  unique times across individuals are $\mathbf{t} = \{0, 1, 2, ...,49\}$, and the total number of unique times is $q = 50$.

In total, there are $n = 101$ patients in the study. Each patient was assigned a clinical label (1-4) denoting their peak severity, with a higher number corresponding to a more severe state. Meanings of labels are defined in Ruffieux et al \cite{ruffieux2023patient} as follows (with corresponding sample size in parentheses): 1, mild symptomatic ($31$); 2, hospitalized without supplemental oxygen ($12$); 3, hospitalized with supplemental oxygen ($19$); and 4, hospitalized with assisted ventilation ($39$). Among those labelled severity level 4, $4$ patients did not survive for the full 7-week window; the exact dates of death are days $16$, $20$, $44$, and $47$ after symptom onset, respectively. 

We apply our proposed dynamic factor analysis approach to the longitudinal (but sparse and irregular) measurements of the $35$ metabolites recorded in the study. As very few patients ($4\%$) died during the study period, we chose to not explicitly model the survival aspect of these data here. Our aim is to discover biological mechanisms and their possible interplay underlying the biological heterogeneity in patient's response to COVID-19 viral infection. Our proposed analysis strategy is described in the next section.

\subsection{Analysis strategy}
\label{data_results}
Ageing and metabolism are inextricably linked, and many changes in metabolites may be related to the ageing process. For example, Singh et al \citep{singh2023taurine} claim that the abundance of the metabolite taurine (measured in this study) decreases during ageing.
To characterise the biological heterogeneity in metabolite dynamics not due to exogenous age but to the response to COVID-19 infection, we regress out age. More precisely, we perform metabolite-specific linear regression, $\tilde{x}_{ijg} = \beta_{0g} + \beta_{1g}a_i + \epsilon_{ijg}$, where $\tilde{x}_{ijg}$ corresponds to the subject and time-specific measurement of the $g$th metabolite, $a_i$ denotes subject-specific age, and $\epsilon_{ijg}$ is the residual error assumed to follow a normal distribution $N(0,\sigma_{g}^2)$; $i=1,\ldots , n$ and $j=1,\ldots , q_i$. We use the measurements from these age-regressed out metabolites (i.e. the residuals) as the observed variables, $x_{ijg}$, in our dynamic factor analysis model; that is, $x_{ijg} = \tilde{x}_{ijg} - \hat{x}_{ijg}$, where $\hat{x}_{ijg}$ is the fitted value of the linear regression.

Our primary analysis corresponds to the fitting of our proposed dynamic factor model for sparse and irregular data  with $q=50$ (corresponding to the set of unique times $\mathbf{t} = \{0, 1, 2, ...,49\}$) and for several different $k$, the number of latent factors, from $2$ to $5$. The value for the tuning parameter, $\lambda$ was obtained through cross-validation. We consider values of $\ln{\lambda}$ between $-4$ and $4$, with step-length $0.5$; resulting in $17$ candidate values of $\lambda$ range from $0.02$ to $54.60$.

We compare the results from our primary analysis with the results from fitting the model of Cai et al \citep{cai2023dynamic}, which assumes zero-mean functions for MOGP, without any penalisation of roughness and implements MCEM for inference on the MOGP hyperparameters. However, for $q=50$, the MCEM implementation failed to return results within $36$ hours. Therefore to allow some level of comparison, we coarsened the set of $50$ unique times $\mathbf{t}$ into a reference/canonical grid, $\mathbf{t}_{\text{ref}}= \{0,7,14,...,49\}$, of $q=8$ unique time points that could be used within the MCEM algorithm. This grid choice was informed by the actual subject-specific observation times, as displayed in Figure~\ref{observed_times}. Moreover, it was motivated by the COVID-19 study design, where in-patients were to be sampled at enrollment then approximately weekly up to 4 weeks from enrollment and thereafter every 2 weeks up to 12 weeks. Out-patients were sampled at enrollment and then approximately 2 and 4 weeks after enrollment. The mapping of the observed times to the times in the reference grid $\mathbf{t}_{\text{ref}}$, where adjacent reference times are 1-week apart, resulted in an approximately constant patient-specific shift across patient's observation times (as shown in Supplementary Figure~1). As the GP we adopt are stationary, their covariance functions will only depend on the time difference rather than specific times. Therefore given this knowledge and that the mapping results in approximately constant patient-specific shifts, there will be minimal impact of the mapping on the estimation of the GP hyperparameters. The specific details of how the mapping to the reference grid was accomplished can be found in Supplementary Section B.1. 
Additionally, we compared the metabolites selected by our approach with those identified in Ruffieux et al \citep{ruffieux2023patient}.

\subsection{Statistical results}
\subsubsection{Primary results under our proposed method}
\label{improvement_after_roughness_penalty}
The primary results using our proposed method with $q = 50$ are presented below. Regardless of the choice of $k$, we consistently identify one factor that can differentiate patients with different clinical severities. Therefore, we chose the simplest model with $k = 2$ as our final model and display results under $k=2$ hereafter. Supplementary Figures 6-7 display results under $k = 3$ as an example of other specifications. 

\begin{figure}[hbt!]
\centering
\includegraphics[width=\textwidth]{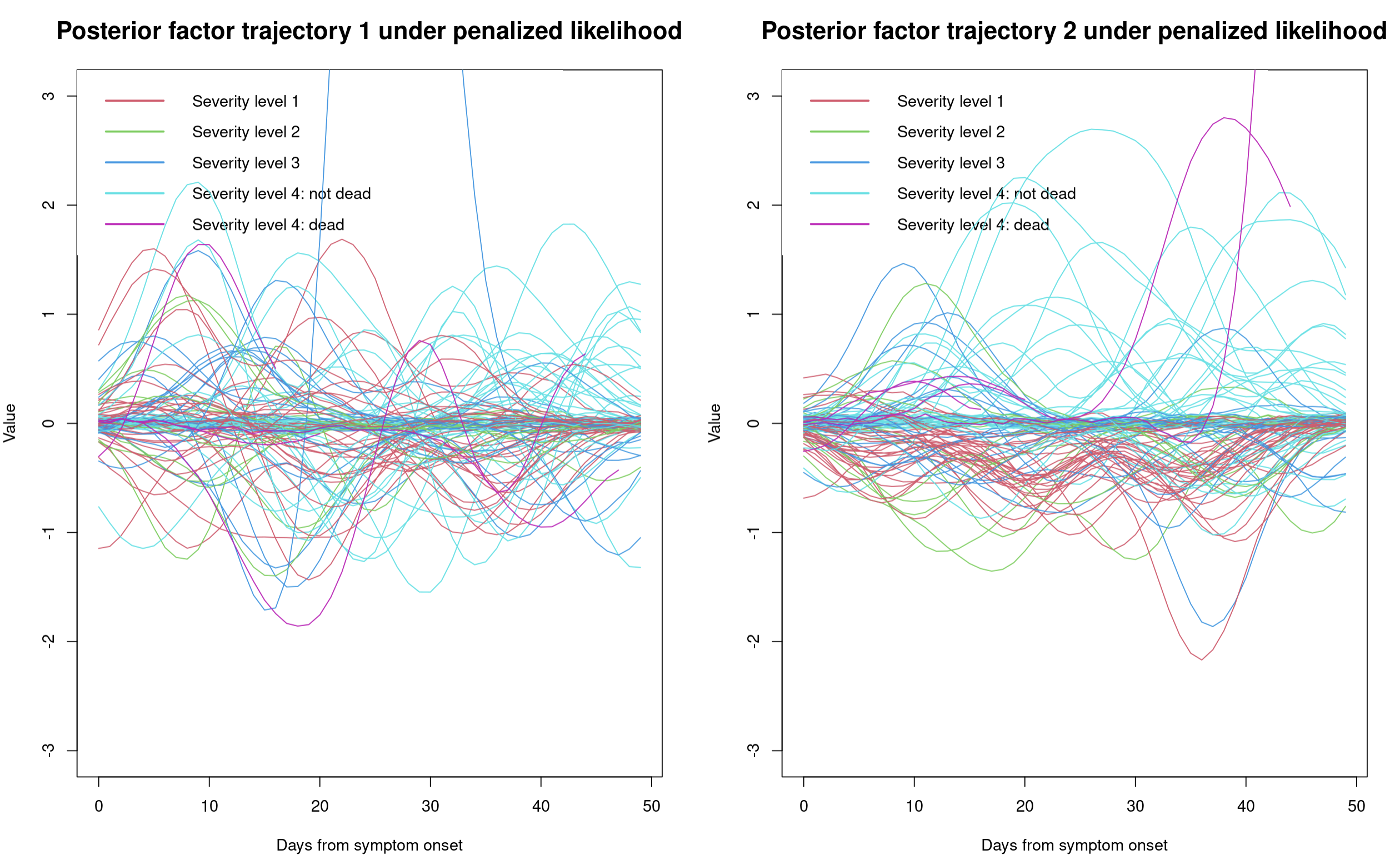}
\caption{Estimated subject-specific posterior factor trajectories (truncated at $y=3$ for clarity; see Supplementary Figure 2 for the full version), using the exact method with $q = 50$. The number of latent factors is pre-specified as $k=2$. For people who did not survive for the full 7-week followup, we plotted their trajectories only before the death date; otherwise trajectories were plotted within a 7-week window after the onset of symptoms.}
\label{posterior_factor_trajectories_q_50_truncated}
\end{figure}

Supplementary Figure 4 indicates that the optimum of $\lambda$ (denoted as $\lambda^{\text{opt}}$) is around $3$, as this value leads to the smallest averaged MAE under cross-validation. Under $\lambda^{\text{opt}}$, the magnitude (i.e. absolute value) of the estimated cross-correlation between factors 1 and 2 is $0.003$, almost negligible. Figure~\ref{posterior_factor_trajectories_q_50_truncated} displays estimated subject-specific factor trajectories. Visually, it appears that the second factor trajectory is able to distinguish patients with the mildest symptoms (i.e., severity level 1) from those with the severest symptoms (i.e., severity level 4) most clearly (See Supplementary Figure 5 for trajectories of patients from severity levels 1 and 4 only). Whereas patients with severity levels 2 and 3 (i.e., intermediate-level severity) could not be differentiated clearly from patients in the two extreme classes.


Moreover, our method is able to estimate the `complete' trajectories of metabolites (i.e. over 7 weeks) via the posterior predictive distribution (described in Supplementary Section A.1). Figure~\ref{biomarker_under_constraint} shows the posterior predictive distribution for the $9$th metabolite (quinolinic acid), which has the largest absolute factor loading on the second factor. We plotted estimated curves of this biomarker for $20$ subjects (randomly selected).

\begin{figure}[htp]
\centering
\includegraphics[width=\textwidth]{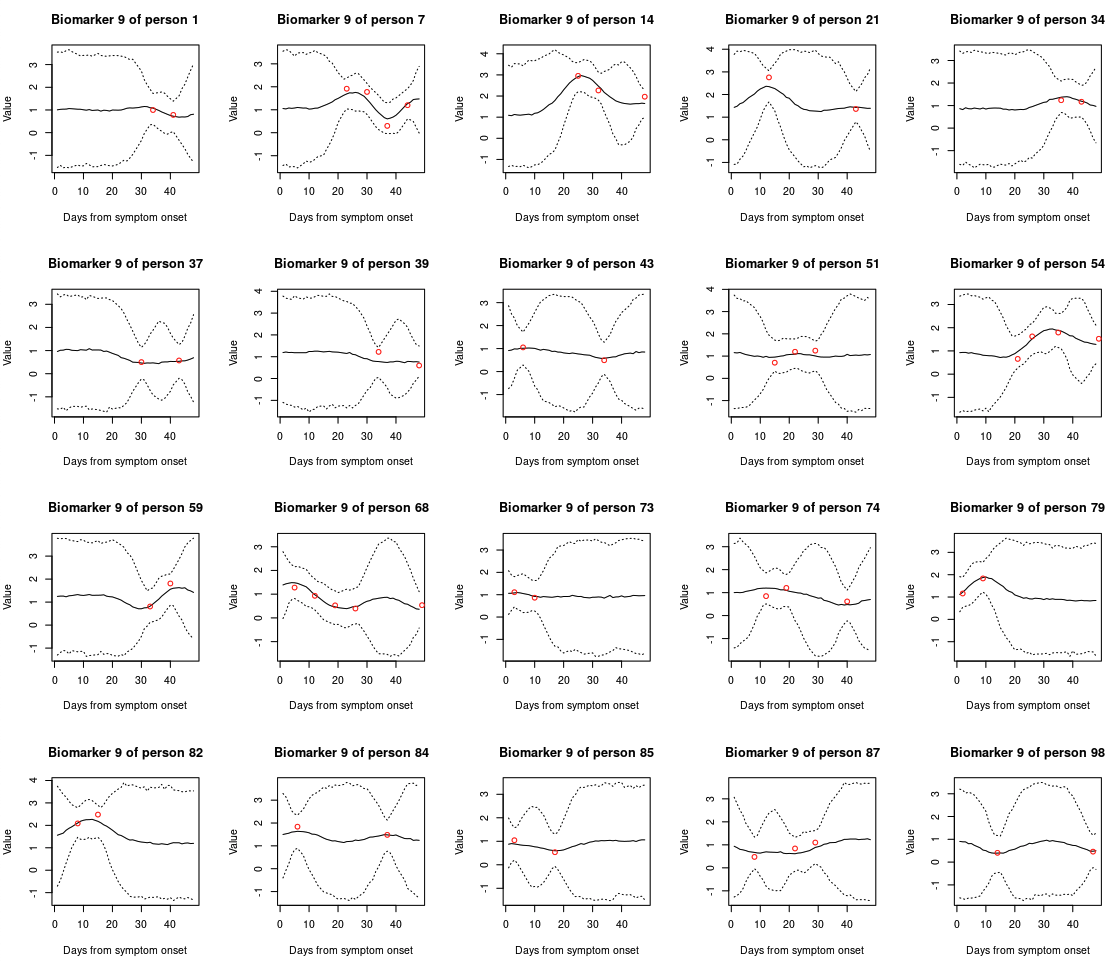}
\caption{Posterior predictive trajectory distributions of the $9$th biomarker for $20$ subjects (randomly selected), under penalized likelihood. Solid and dotted lines represent point estimate and $95\%$ credible intervals, respectively. Red points denote observed measurements.}
\label{biomarker_under_constraint}
\end{figure}

\subsubsection{Results under the comparator method}
\label{overfitting}
\begin{figure}[htp]
\centering
\includegraphics[width=\textwidth]{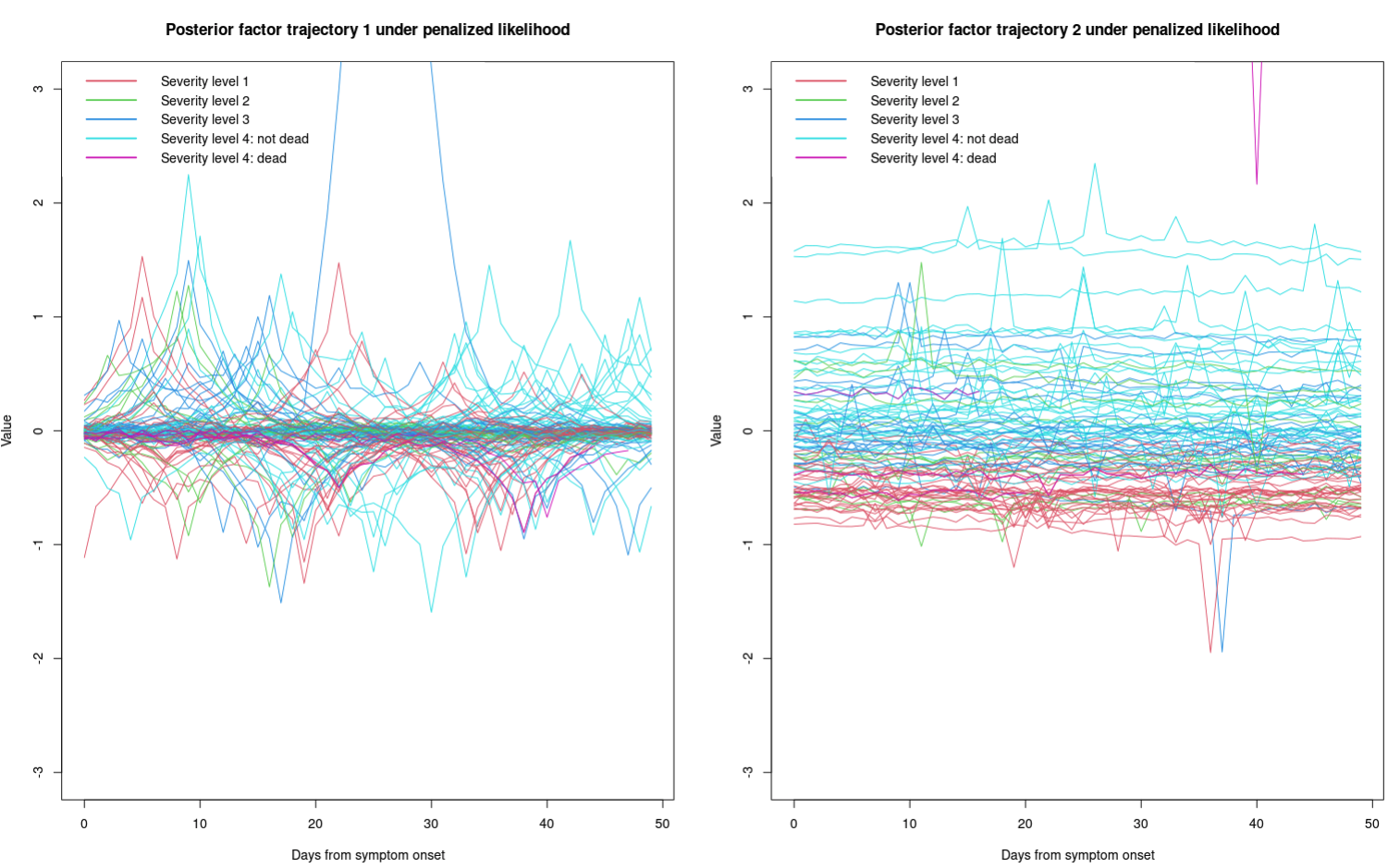}
\caption{Estimated subject-specific posterior factor trajectories using the comparator method \citep{cai2023dynamic} with $q = 8$ (truncated at $y=3$ for clarity). The number of latent factors is pre-specified as $k=2$. For people who did not survive for the full 49-day followup, we plotted their trajectories only before the death date; otherwise trajectories were plotted within a 7-week window after the onset of symptoms.}
\label{factor_no_constraint}
\end{figure}

\begin{figure}[htp]
\centering
\includegraphics[width=\textwidth]{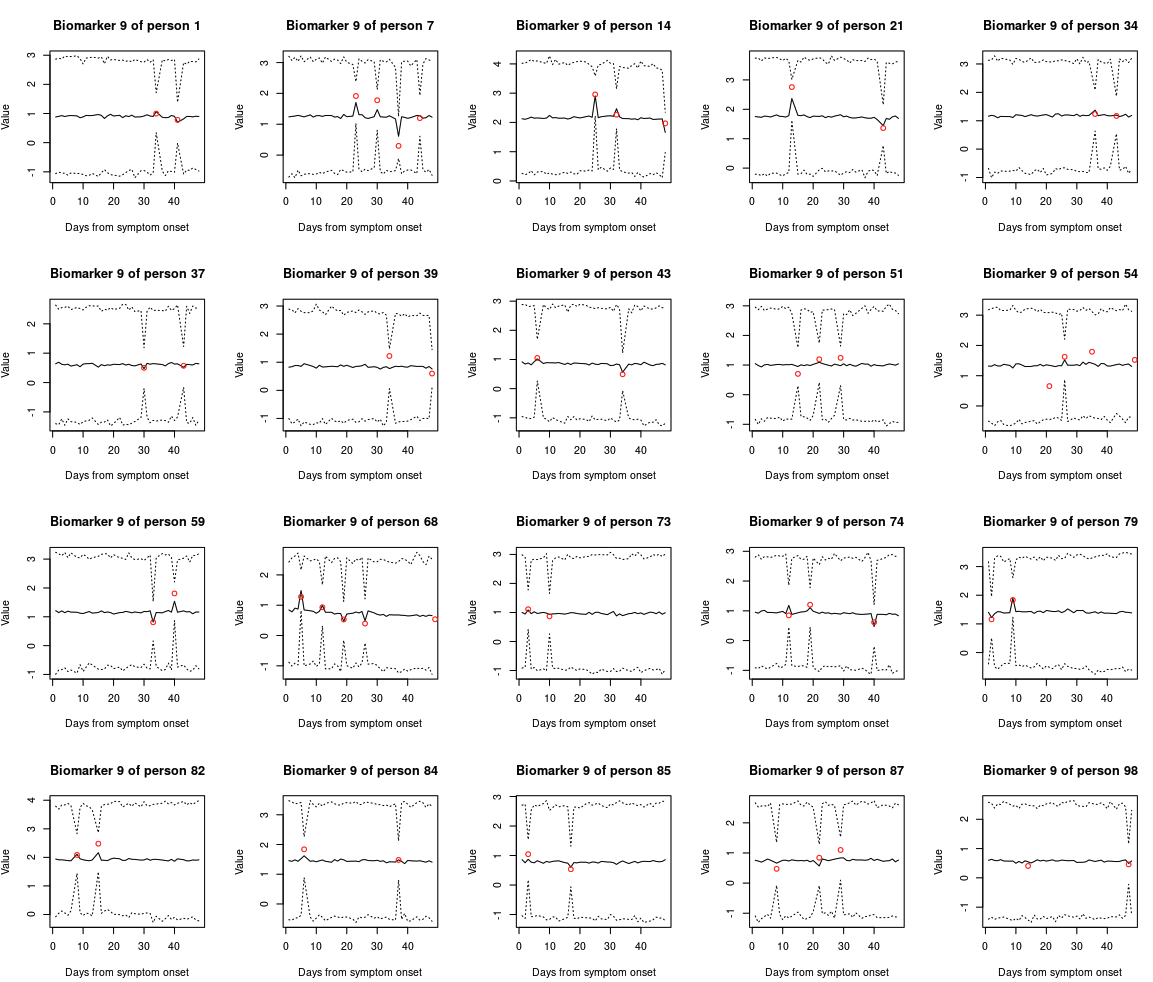}
\caption{Posterior predictive trajectory distribution of the $9$th biomarker for $20$ subjects (randomly selected), by applying the model in Cai et al \citep{cai2023dynamic} to the motivating example. Solid and dotted lines represent point estimate and $95\%$ credible intervals, respectively. Red points denote observed measurements.}
\label{no_constraint}
\end{figure}

When $k = 2$, the comparator method could not return estimates within $36$ hours for the $q = 50$ unique observation times. To speed up its computation, we implemented the reference grid mapping strategy with $q = 8$. Figure~\ref{factor_no_constraint} displays estimated pathway trajectories, which are implausibly spiky. This reflects little correlation between adjacent time points, and it is implausible biologically. In contrast, their counterparts in Figure~\ref{posterior_factor_trajectories_q_50_truncated} (under our model) are much smoother. Moreover, the poor performance in recovering latent factor trajectories results in poor performance in the estimation of biomarker trajectories (Figure~\ref{no_constraint}). These curves are almost flat, and only change towards measured values at observed time points, reflecting severe overfitting. In contrast, the estimated biomarker trajectories under our method display a more plausible temporal trend (Figure~\ref{biomarker_under_constraint}). 



Sparse observations may be the cause of the comparator model failing to return meaningful temporal trends. The model utilized GP to describe the trajectories of pathways, and the flexibility of GP tends to lead to overfitting when the number of measurements is few. If no regularization is imposed, results are mostly implausible. On the contrary, our model can handle sparse data well, due to the constraints we introduced in Section~\ref{handle_sparse}. 

\subsection{Biological interpretation}
\label{bio_interpretation}
To interpret the factor loading results biologically, we used the online bioinformatics platform \href{https://www.metaboanalyst.ca/MetaboAnalyst/Secure/pathway/PathResultView.xhtml}{MetaboAnalyst 5.0} to perform pathway enrichment analysis for the top $10$ metabolites with the highest absolute loading on the second factor.  All $35$ metabolites measured in the COVID-19 study were inputted as the background/reference. 

\begin{figure}[htp]
\centering
\includegraphics[width=\textwidth]
{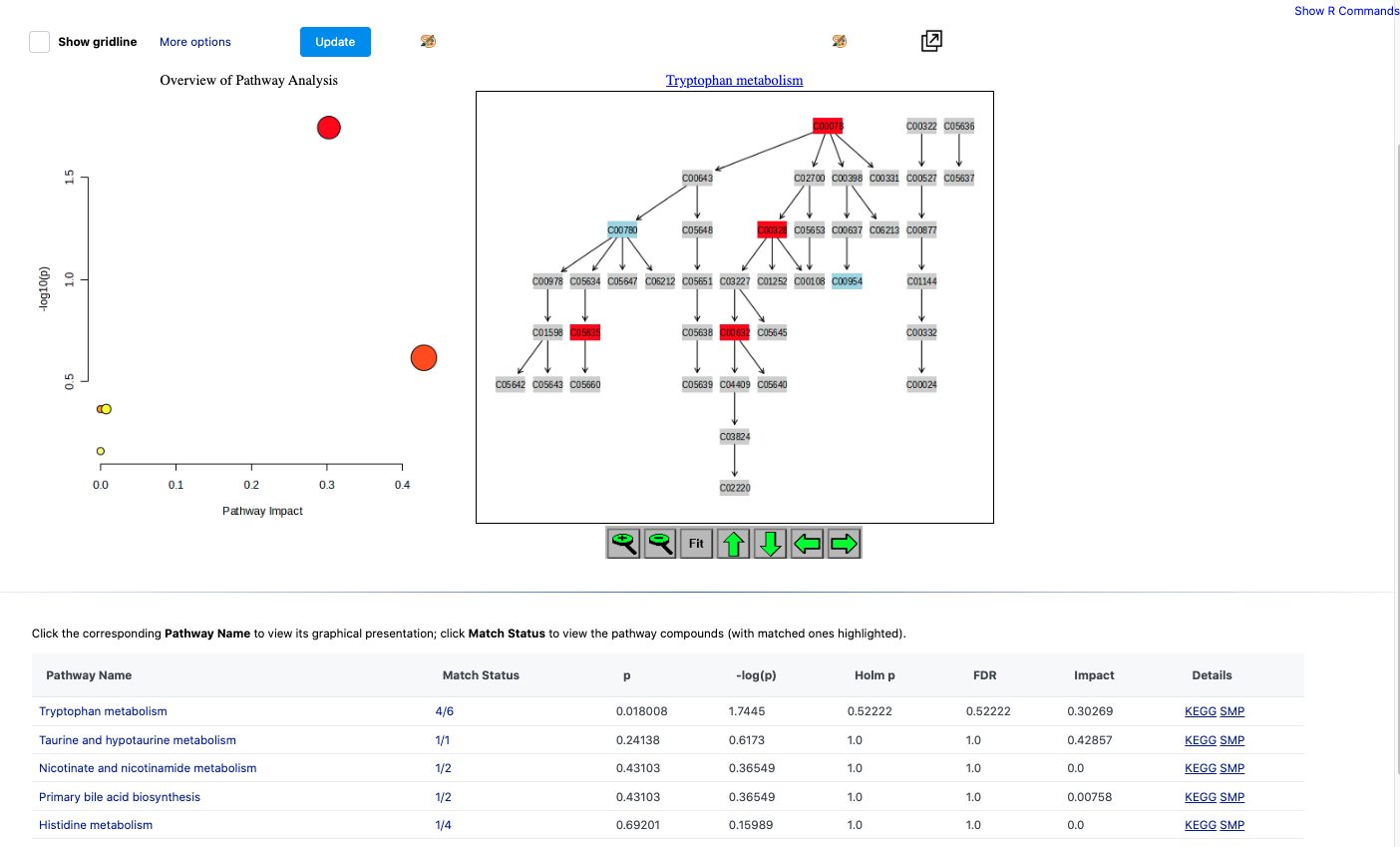}
\caption{Pathway analysis results by the online bioinformatics platform \href{https://www.metaboanalyst.ca/MetaboAnalyst/Secure/pathway/PathResultView.xhtml}{MetaboAnalyst 5.0}. The top left panel shows an overview of pathway analysis, with y-axis denoting p-value of pathway enrichment analysis and x-axis representing impact of the selected metabolites on the pathway. The top right panel shows details of the enriched pathway: colored (i.e., not grey) elements represent metabolites in this pathway that are uploaded as `background', with red and blue elements denoting `selected' and `unselected' metabolites by our method, respectively. The bottom panel supplies numerical details for the top figures.}
\label{metaboanalyst}
\end{figure}

Figure~\ref{metaboanalyst} displays the result. 
The enrichment analysis returns a p-value of $0.018$ for the tryptophan metabolism pathway (see column `p' in the bottom panel or through the y-axis corresponding to `-log(p)' in the top left panel), suggesting selected metabolites are significantly enriched in this pathway. Its specific mechanism is provided in the top right panel. The importance of this pathway in COVID-19 has been previously established. Tryptophan metabolism via the kynurenine pathway has been found to be the top pathway influenced by COVID-19 infection, and it has been argued that this should be the focus of COVID-19 immunotherapy \citep{badawy2023kynurenine}. Furthermore, the pathway topology analysis, which evaluates the impact of the chosen metabolites on the pathway, returns a value of $0.30$ (the column `impact' in the bottom panel) for tryptophan metabolism pathway. This number indicates the selected metabolites are in key positions of this pathway. 

In addition, all metabolites identified as signatures of COVID-19 patients in Ruffieux et al \cite{ruffieux2023patient} (introduced in Section~\ref{background}) have been selected by our method. Furthermore, we identified a novel target `taurine', which has been receiving increased attention clinically \citep{iwegbulem2021role} yet was overlooked in the previous analysis \citep{ruffieux2023patient}. Our analysis suggests that a higher level of taurine is associated with milder symptoms, which is consistent with existing knowledge; for example, van Eijk et al claims that `taurine should be regarded as a promising supplementary therapeutic option in COVID-19' \citep{van2022disease}. In addition, the association between taurine and the kynurenine pathway, characterized by the corresponding factor loading estimate, would be a lot weaker if age was not regressed out. The median estimate of the factor loading after adjusting for age is $-0.25$ ($95\%$ credible interval $(-0.33, -0.17)$), whereas it is $-0.09$ ($95\%$ credible interval $(-0.21, 0)$) without age adjustment. Adjusting for age (e.g. known to affect taurine level from prior biological knowledge \citep{singh2023taurine}) has resulted in a refinement of the factor loading structure, leading to increased specificity in the identification of metabolites instrumental in the underlying mechanisms of COVID-19 disease dynamics.


\section{Discussion}
\label{discussion}
In this paper, we propose a dynamic factor analysis approach that can uncover latent biological structure and estimate the 'complete' trajectories of biomarkers, from sparsely and irregularly collected longitudinal biomarker data in studies with relatively large sample size. Specifically, we introduce a penalized MOGP with non-zero mean functions to model the latent factor trajectories, which not only accounts for the cross-correlations between factors, but also mitigates against overfitting caused by sparse observations. For inference on MOGP hyperparameters, we develop an StEM algorithm that scales well with 'large' sample size. Our method is implemented in our R package DFA4SIL. It can be viewed as an extension to the model proposed in Cai et al \citep{cai2023dynamic}, which works well in small studies with frequent measurements. We anticipate this extension will facilitate the use of dynamic factor analysis with potential interactions between factors in a wider range of biomedical data applications. Application to the metabolite data in a COVID-19 study demonstrates the ability of our model to identify and characterise essential biological pathways (and corresponding instrumental metabolites) during disease progression and to estimate/predict the  trajectories of metabolites over time after infection.

Although the cross-correlation in the COVID-19 metabolite data application is estimated to be close to zero (indicating the independence of the two latent factors), non-zero estimates are possible with our method if factors are truly correlated, which has been shown in the simulation study. Theoretically, it is preferable to apply a model that allows for dependence among factors (as is proposed here), rather than a model that assumes independence {\em{a priori}}. The data would then allow us to determine whether or not independence amongst factors is appropriate. Moreover, our method is applicable to multiple biomarker types and able to capture the potential cross-correlations among different types. As multiple data types can capture a wider range of biological mechanisms that are in play after virus infection, further insight may arise in this setting.

An alternative approach to control the roughness of GP is to directly limit the support of key hyperparameters. For instance, a bounded length-scale has been implemented for squared exponential and Mat\'{e}rn covariance functions, based on the spectral density \citep{topa2012gaussian}. However, it is unclear how to extend this approach to complex covariance functions, such as the covariance function induced by the convolution process framework we introduced in Section~\ref{cp}.

In the future, there are several research directions worth pursuing. First, the computational complexity of the StEM algorithm used to determine the MLE of MOGP hyperparameters scales quadratically with $kq$. When $kq$ is large (e.g., when observed times are highly irregular across individuals or when the pre-specified number of latent factors is large), approximation methods may be used to further speed up the computation if necessary. For example, a reference grid approach could be implemented along the lines as was done in the real-data application when applying the original method of Cai et al. \citep{cai2023dynamic} to the COVID-19 data. However, some careful consideration should be given to what is an appropriate reference grid to choose which balances computational speed gains with the resulting loss of information from coarsening and allows the coarsening mechanism to be ignored in the Bayesian analysis.
Alternatively, approximation methods based on variational inducing kernels, as proposed by Alvarez et al \cite{alvarez2011computationally}, or alternative algorithms such as variational inference \citep{titsias2009variational, alvarez2010efficient, hensman2013gaussian} may be propitious.

Moreover, the current model in Equation~\ref{integrated_model} does not take the clinical outcome information into account. Instead, we first fit an unsupervised model that does not use clinical labels, then check if there is a difference in the estimated factor trajectories of people with different levels of clinical severity by visual inspection. A model that jointly models the underlying factor structure and the relationship between factors and outcomes would be of future interest, as previous researchers have found that an outcome-guided model can improve the estimation of the underlying factor structure (e.g. better estimation of the factor loading matrix) \citep{li2016supervised, tu2019forecasting} and lead to the discovery of more clinically relevant structure. 


 \section{Acknowledgement}
We thank NIHR BioResource volunteers for their participation, and gratefully acknowledge NIHR BioResource centres, NHS Trusts and staff for their contribution. We thank the National Institute for Health and Care Research, NHS Blood and Transplant, and Health Data Research UK as part of the Digital Innovation Hub Programme. We acknowledge support from the NIHR Cambridge Biomedical Research Centre (NIHR203312). The views expressed are those of the author(s) and not necessarily those of the NHS, the NIHR or the Department of Health and Social Care. We are also grateful to Hélène Ruffieux for sharing the data; Christoph Hess and Joy Edwards-Hicks for the helpful discussion on interpreting the results. For the purpose of open access, the author has applied a Creative Commons Attribution (CC BY) licence to any Author Accepted Manuscript version arising.

\section{Funding information}
This work is supported through the United Kingdom Medical Research Council programme grants \texttt{MC\_UU\_00002/2}, \texttt{MC\_UU\_00002/20}, \texttt{MC\_UU\_00040/02} and \texttt{MC\_UU\_00040/04}.

\section{Data availability statement}
The data are from the ``CITIID-NIHR BioResource COVID-19 Collaboration''. It is available at \url{https://www.covid19cellatlas.org/patient/citiid/} (clinical data) and \url{https://doi.org/10.5281/zenodo.7277164} (metabolite data).

{\it Conflict of Interest}: None declared.

\bibliography{main}

\clearpage

\renewcommand*{\algorithmcfname}{Supplementary Algorithm}

\renewcommand{\figurename}{Supplementary Figure}
\renewcommand{\tablename}{Supplementary Table}

\setcounter{figure}{0} 
\setcounter{table}{0} 

\counterwithout{equation}{section}

\clearpage

{\centering{\huge {\textbf {Supplementary Materials}} }}

\section*{A. Mathematical derivations}
\subsection*{A.1 Full conditionals of the Gibbs sampler for the proposed model}


Throughout the following derivation, ``$\circ$'' denotes element-wise multiplication, ``$-$'' denotes observed data and all parameters in the model other than the parameter under derivation, ``$||\mathbf{z}||^2$'' denotes the sum of squares of each element of the vector $\mathbf{z}$, ``$\text{diag}(\mathbf{z})$'' denotes a diagonal matrix with the vector $\mathbf{z}$ as its main diagonal elements, ``$\text{MVN}\left(\mathbf{z}; \boldsymbol{\mu}_{\mathbf{z}}, \Sigma_{\mathbf{z}}\right)$'' denotes the probability density function of a multivariate normal distribution with mean $\boldsymbol{\mu}_{\mathbf{z}},$ and variance $\Sigma_{\mathbf{z}}$, and similar interpretations apply to other distributions. ``$\mathbf{z}^{T}$'' denotes the transpose of the vector or matrix $\mathbf{z}$, and ``pos'' is short for `posterior probability'.

\begin{itemize}
    \item Full conditional for the latent factors $\vect({\mathbf{Y}_{i,\text{aug}}}^{T}),i=1,\ldots,n$
    \begin{align*}
    \begin{split}
        f(\vect({\mathbf{Y}_{i,\text{aug}}}^{T}) \mid -)
        & =  f(\vect({\mathbf{Y}_{i}^{T}}) \mid -) \cdot     f(\vect({\mathbf{Y}_{i,\text{add}}^{T}}) \mid \vect({\mathbf{Y}_{i}^{T}}),\Sigma_{\mathbf{Y}_{\text{aug}}}(\boldsymbol{\Theta}, \mathbf{t}))\\
    \end{split}
    \end{align*}
    We first sample $\vect({\mathbf{Y}_{i}^{T}})$ according to:
    \begin{align*}
    \begin{split}
f(\vect({\mathbf{Y}_{i}^{T}})\mid -)
        & \propto \text{MVN} \left(\vect({\mathbf{X}_i^{T}}); \  \vect({\mathbf{M}_i^{T}})+ \mathbf{L}_i^{*} \vect({\mathbf{Y}_{i}^{T}}), \ \Sigma_{\mathbf{X}_i} \right)
        \cdot \text{MVN}\left(\vect({\mathbf{Y}_{i}^{T}}); \ \mathbf{0},\ \Sigma_{\mathbf{Y}_i}(\boldsymbol{\Theta}, \mathbf{t}_i)\right) \\
    &=\text{MVN}\left(\mu_{\mathbf{Y}_{i,}}^{\text{pos}}, \ \Sigma_{\mathbf{Y}_{i, }}^{\text{pos}}\right),
    \end{split}
    \end{align*}
     with
    \begin{align*}
      \begin{split}
        \Sigma^{\text{pos}}_{\mathbf{Y}_{i,}}&=[{\mathbf{L}_i^{*}}^{T}\Sigma_{\mathbf{X}_{i}}^{-1}\mathbf{L}_i^{*}+\Sigma_{\mathbf{Y}_i}(\boldsymbol{\Theta}, \mathbf{t}_i)^{-1}]^{-1} \\
    \mu^{\text{pos}}_{\mathbf{Y}_{i,}}&=\Sigma^{\text{pos}}_{\mathbf{Y}_{i,}} ({\mathbf{L}_i^{*}}^{T}\Sigma_{\mathbf{X}_{i}}^{-1}(\vect({\mathbf{X}_i^{T}})-\vect({\mathbf{M}_i^{T}}))).
      \end{split}
    \end{align*}
    Then we sample $\vect({\mathbf{Y}_{i,\text{add}}^{T}})$ from $f(\vect({\mathbf{Y}_{i,\text{add}}^{T}})|\vect({\mathbf{Y}_{i}^{T}}),\ \Sigma_{\mathbf{Y}_{\text{aug}}}(\boldsymbol{\Theta}, \mathbf{t}))$, a MVN distribution due to the property of the MOGP model \citep{shi2011gaussian}. \\
    
    In the above equations: 
    
    \begin{itemize}
    
     \item  $\Sigma_{\mathbf{Y}_{\text{aug}}}(\boldsymbol{\Theta}, \mathbf{t})$ is the covariance matrix of factor scores at full time $\mathbf{t}$, and $\Sigma_{\mathbf{Y}_i}(\boldsymbol{\Theta}, \mathbf{t}_i)$ is a sub-matrix of it (at subject-specific time points);
     
     \item  
    $\mathbf{L}_i^{*}$ is constructed using components of the factor loading matrix $\mathbf{L}$:  $\mathbf{L}_i^{*}=(\mathbf{L}_1^{*},...,\mathbf{L}_p^{*})^{T} \in \mathbb{R}^{pq_i \times kq_i}$,  where $(\mathbf{L}_g^{*})^{T}=({\text{diag}(l_{g1})}_{q_i \times q_i},...,{\text{diag}(l_{gk})}_{q_i \times q_i}) \in \mathbb{R}^{q_i \times kq_i}$; 
    
    \item 
    
    $\Sigma_{\mathbf{X}_i}=\text{diag}(({\phi_1^2})_{\times q_i},...,({\phi_p^2})_{\times q_i}) \in \mathbb{R}^{pq_i \times pq_i}$, where $\ ({\phi_a^2})_{\times q_i}$ represents a $q_i$-dimensional row vector consisting of the scalar ${\phi_a^2}$.
    
    \end{itemize}
    
    Note that when coding the algorithm, there is no need to directly create the $pq_i \times pq_i$ diagonal matrix $\Sigma_{\mathbf{X}_{i}}^{-1}$ as the memory will be exhausted. To calculate the term ${\mathbf{L}_i^{*}}^{T}\Sigma_{\mathbf{X}_{i}}^{-1}$, we can use the property of multiplication of a diagonal matrix: post-multiplying a diagonal matrix is equivalent to multiplying each column of the first matrix by corresponding elements in the diagonal matrix. \\
    
           
           %
           
          \item Full conditional for the binary matrix $\mathbf{Z}$\\ 
          Let $\mathbf{Z}_{g \cdot} = (Z_{g1},...,Z_{gk})$ denote the $g$th row of the matrix $\mathbf{Z}, \ g=1,...,p$; then, 
          \begin{align*}
          \begin{split}
               f(\mathbf{Z}_{g \cdot}|-) 
               & \propto \prod_{i=1}^{n}\text{MVN} \left(\mathbf{x}_{ig}; \ \boldsymbol{\mu}_{ig} + (\mathbf{A}_{g \cdot} \circ \mathbf{Z}_{g \cdot})\mathbf{Y}_{i}, \ \text{diag}(\phi_g^{2}, \ q_i) \right) \cdot \prod_{a=1}^{k} \text{Bernoulli}(Z_{ga}; \pi_a),
          \end{split}
          \end{align*}
          We calculate the posterior probability under $2^k$ possible values of $\mathbf{Z}_{g \cdot}$ based on the above formula, then sample with corresponding probability.
          
          
          
          \item Full conditional for the regression coefficient matrix $\mathbf{A}$ \\
          Let $\mathbf{A}_{g \cdot} = (A_{g1},...,A_{gk})$ denote the $g$th row of the matrix $\mathbf{A}, \ g=1,...,p$; then, 
          \begin{align*}
          \begin{split}
               f(\mathbf{A}_{g \cdot}|-) 
               & \propto \prod_{i=1}^{n}\text{MVN}\left(\mathbf{x}_{ig}; \ \boldsymbol{\mu}_{ig} + (\mathbf{A}_{g \cdot} \circ \mathbf{Z}_{g \cdot})\mathbf{Y}_{i}, \ \text{diag}(\phi_g^{2}, \ q_i)\right) \cdot \text{MVN}\left(\mathbf{A}_{g \cdot}; \ \mathbf{0},\ \text{diag}(\boldsymbol{\rho}^2)\right) \\
               & =\text{MVN}\left(\mathbf{A}_{g \cdot}; \ \mu_{\mathbf{A}_{g \cdot}}^{\text{pos}}, \ \Sigma_{\mathbf{A}_{g \cdot}}^{\text{pos}}\right),
          \end{split}
          \end{align*}
         
          where 
          \begin{align*}
              \begin{split}
              \Sigma_{\mathbf{A}_{g \cdot}}^{\text{pos}}&= \left({\frac{\text{diag}(\mathbf{Z}_{g\cdot})(\sum_{i=1}^{n}{\mathbf{Y}_{i}}^{T}\mathbf{Y}_{i})\text{diag}(\mathbf{Z}_{g\cdot})}{\phi^2_g}+\text{diag}(\frac{1}{\boldsymbol{\rho^2}})}\right)^{-1} \\
              \mu_{\mathbf{A}_{g \cdot}}^{\text{pos}}&=\frac{\Sigma_{\mathbf{A}_{g \cdot}}(\text{diag}(\mathbf{Z}_{g\cdot})\sum_{i=1}^{n}{\mathbf{Y}_{i}}(\mathbf{x}_{ig}-\boldsymbol{\mu}_{ig}))}{\phi^2_g}.
              \end{split}
          \end{align*}
          and $\boldsymbol{\rho^2}=(\rho^2_1,...,\rho^2_a)$.

          \item Full conditional for the intercept $\mu_{ig}, i=1,...,n; g=1,...,p$
          \begin{align*}
          \begin{split}
              f(\mu_{ig}|-)
              &\propto \prod_{j=1}^{q_i} \text{N}(x_{ijg}; \mu_{ig} + \sum_{a=1}^{k}l_{ga}y_{ija}, \phi^2_g) \cdot \text{N}(\mu_{ig};\mu_g, \sigma^2_g) \\
              & = \text{N}(\mu_{ig}; \mu_{ig}^{\text{pos}}, \ \sigma_{ig}^{2,\text{pos}}),
          \end{split}
          \end{align*}
          where 
          \begin{align*}
          \begin{split}
             \sigma_{ig}^{2,\text{pos}} & = {(\frac{1}{\sigma^2_g} + \frac{q_i}{\phi^2_g})}^{-1} \\
             \mu_{ig}^{\text{pos}} & = \left(\frac{\mu_g}{\sigma^2_g} + \frac{\sum_{j=1}^{q_i}(x_{ijg}-\sum_{a=1}^{k}l_{ga}y_{ija})}{\phi^2_g}\right) \cdot \sigma_{ig}^{2,\text{pos}}
          \end{split}
          \end{align*}
         
          \item Full conditional for $\pi_a,\ a=1,...,k$ 
          \begin{align*}
              \begin{split}
              f(\pi_a|-)
              & \propto \prod_{a=1}^{k}\text{Bernoulli}(Z_{ga}; \pi_a) \cdot \text{Beta}(\pi_a; c_0, d_0)\\
              &=\text{Beta}\left(c_0+\sum_{g=1}^{p}Z_{ga},\ d_0+\sum_{g=1}^{p}(1-Z_{ga})\right) \\
              \end{split}
          \end{align*}
          
          \item Full conditional for $\rho_a^2,\ a=1,...,k$ 
           \begin{align*}
              \begin{split}
                f(\rho_a^2|-)
                & \propto \prod_{g=1}^{p} \text{N}(A_{ga};0, \rho^2_a) \cdot \text{Inverse-Gamma}(\rho_a^2; c_1, d_1) \\
                & =\text{Inverse-Gamma} \left(c_1+\frac{p}{2},\ d_1+\frac{1}{2}\sum_{g=1}^{p}A_{ga}^2 \right) \\
              \end{split}
          \end{align*}
          
          \item Full conditional for $\sigma_g^2,\ g=1,...,p$
          \begin{align*}
          \begin{split}
            f(\sigma_g^2|-) 
            & \propto \prod_{i=1}^{n} \text{N}(\mu_{ig};\mu_g, \sigma^2_g) \cdot \text{Inverse-Gamma}(\sigma^2_g; c_2, d_2) \\
            &=\text{Inverse-Gamma} \left (c_2+\frac{1}{2}n, \ d_2+\frac{1}{2}\sum_{i=1}^{n}{(\mu_{ig}-\mu_g})^2 \right)
          \end{split}
          \end{align*}

          \item Full conditional for $\phi_g^2,\ g=1,...,p$
          \begin{align*}
          \begin{split}
            f(\phi_g^2|-) 
            & \propto \prod_{i=1}^{n}\text{MVN}(\mathbf{x}_{ig}; \ \boldsymbol{\mu}_{ig} + (\mathbf{A}_{g \cdot} \circ \mathbf{Z}_{g \cdot})\mathbf{Y}_{i}, \ \text{diag}(\phi_g^{2}, \ q_i)) \cdot \text{Inverse-Gamma}(\phi^2_g; c_3, d_3) \\
            &=\text{Inverse-Gamma}\left(c_3+\frac{1}{2}\sum_{i=1}^{n} q_{i}, \ d_3+\frac{1}{2}\sum_{i=1}^{n}{||\mathbf{x}_{ig}-\boldsymbol{\mu}_{ig}-(\mathbf{A}_{g\cdot} \circ \mathbf{Z}_{g\cdot})\mathbf{Y}_i||}^2 \right)
          \end{split}
          \end{align*}
           

           \item Full conditional for predictions of biomarker expression (only implemented when recovering the full trajectories of biomarkers, during the Gibbs-after-StEM stage)
           
           Suppose that $\mathbf{X}_i^{\text{new}}, \mathbf{Y}_i^{\text{new}}$ represent predicted biomarker expression and factor expression of the $i$th individual at new time points, respectively. The posterior predictive distribution under StEM-algorithm-returned $\widehat{\boldsymbol{\Theta}}^{\text{MLE}}$ can be expressed as 
\begin{align*}
\begin{split}
    f(\mathbf{X}_i^{\text{new}}|\ \widehat{\boldsymbol{\Theta}}^{\text{MLE}}, \boldsymbol{\Omega}_{\text{aug}}) 
    & = \int f(\mathbf{X}_i^{\text{new}}, \mathbf{Y}_i^{\text{new}}|\ \widehat{\boldsymbol{\Theta}}^{\text{MLE}}, \boldsymbol{\Omega}_{\text{aug}}) d \mathbf{Y}_i^{\text{new}} \\
    & = \int f(\mathbf{X}_i^{\text{new}}|\mathbf{Y}_i^{\text{new}}, \boldsymbol{\Omega}_{\text{aug}}) \cdot f(\mathbf{Y}_i^{\text{new}}| \widehat{\boldsymbol{\Theta}}^{\text{MLE}},\mathbf{Y}_{i}) d \mathbf{Y}_i^{\text{new}},
\end{split}
\end{align*}
where the first term of the integrand is a MVN because of the assumed factor model, and the second term is also a MVN because of the assumed DGP model on latent factor trajectories \citep{shi2011gaussian}. Therefore, once a sample of parameters $\boldsymbol{\Omega}_{\text{aug}}$ is generated, the sample of ${\mathbf{Y}_i^{\text{new}}}$ can be generated from $f(\mathbf{Y}_i^{\text{new}}| \ \widehat{\boldsymbol{\Theta}}^{\text{MLE}},\mathbf{Y}_{i})$, then the sample of ${\mathbf{X}_i^{\text{new}}}$ can be generated from $f(\mathbf{X}_i^{\text{new}}|\ {\mathbf{Y}_i^{\text{new}}},\boldsymbol{\Omega})$.

\end{itemize}

\subsection*{A.2 Identifiability of the covariance matrix for latent factors}
To facilitate illustrating the identifiability issue, we first re-express the proposed model in Section 2.3 as,
\begin{equation}
\begin{split}
\vect({\mathbf{X}_i^{T}})&=\vect({\mathbf{M}_i^{T}})+\mathbf{L}_i^{*} \vect({\mathbf{Y}_{i}^{T}})+ \vect({\mathbf{E}_i^{T}}), \\
 \vect({\mathbf{Y}_{i}}^{T}) &\sim \text{MVN}(\mathbf{0},\Sigma_{\mathbf{Y}_i}), \\
\vect({\mathbf{E}_i^{T}}) &\sim \text{MVN}(\mathbf{0},\Sigma_{\mathbf{X}_i}),\\
\end{split}
\end{equation}
where $\vect({\mathbf{X}_i^{T}})$, $\vect({\mathbf{M}_i^{T}})$, and $\vect({\mathbf{E}_i^{T}})$ are vectorized from matrices $\mathbf{X}_{i}^{T}$, $\mathbf{M}_{i}^{T}$, and $\mathbf{E}_{i}^{T}$, respectively. To make the above equations hold, $\mathbf{L}_i^{*}$ and $\Sigma_{\mathbf{X}_i}$ are constructed using components of $\mathbf{L}$ and $\boldsymbol{\phi}$, respectively. Specifically, 
\begin{itemize}  

\item $\mathbf{L}_i^{*}=(\mathbf{L}_1^{*},...,\mathbf{L}_p^{*})^{T} \in \mathbb{R}^{pq_i \times kq_i}$,  where $(\mathbf{L}_g^{*})^{T}=({\text{diag}(l_{g1})}_{q_i \times q_i},...,{\text{diag}(l_{gk})}_{q_i \times q_i}) \in \mathbb{R}^{q_i \times kq_i}$; 

\item $\Sigma_{\mathbf{X}_i}=\text{diag}(({\phi_1^2})_{1\times q_i},...,({\phi_p^2})_{1\times q_i}) \in \mathbb{R}^{pq_i \times pq_i}$, where $\ ({\phi_a^2})_{1\times q_i}$ represents a $q_i$-dimensional row vector consisting of the scalar ${\phi_a^2}$.
    
\end{itemize}

The distribution of $\vect({\mathbf{X}_i^{T}})$ after integrating out $\vect({\mathbf{Y}_{i}^{T}})$, is 
\begin{equation}
\label{marginal_distribution}
\begin{split}
\vect({\mathbf{X}_i^{T}}) \ | \ \vect({\mathbf{M}_i^{T}}), \mathbf{L}_i^{*}, \Sigma_{\mathbf{X}_i}, \Sigma_{\mathbf{Y}} \sim \text{MVN}(\vect({\mathbf{M}_i^{T}}), \mathbf{L}_i^{*}\Sigma_{\mathbf{Y}_{i}}(\mathbf{L}_i^{*})^{T}+\Sigma_{\mathbf{X}_i}),
\end{split}
\end{equation}
where $\Sigma_{\mathbf{Y}_{i}}$ is a sub-matrix of $\Sigma_{\mathbf{Y}}$ that characterizes the covariance structure for $\vect({\mathbf{Y}_{i}^{T}})$. 

As noted, there is an identifiability issue with the covariance matrix $\Sigma_{\mathbf{Y}_{i}}$. This issue arises from the invariance of the covariance $\mathbf{L}_i^{*}\Sigma_{\mathbf{Y}_{i}}(\mathbf{L}_i^{*})^{T}+\Sigma_{\mathbf{X}_i}$ in Equation \ref{marginal_distribution}. The uniqueness of $\Sigma_{\mathbf{X}_i}$ has been ensured in previous research \citep{ledermann1937rank, bekker1997generic, conti2014bayesian, papastamoulis2022identifiability}; given its identifiability, we are concerned with identifiability of $\Sigma_{\mathbf{Y}_{i}}$. Non-identifiability is present because for any non-singular transformation matrix $\mathbf{D} \in \mathbb{R}^{kq_i \times kq_i}$, the expression $\mathbf{L}_i^{*}\Sigma_{\mathbf{Y}_{i}}(\mathbf{L}_i^{*})^{T}+\Sigma_{\mathbf{X}_i}$ are equal under these two sets of estimators for $\mathbf{L}_i^{*}$,  $\vect({\mathbf{Y}_{i}^{T}})$ and $\Sigma_{\mathbf{Y}_{i}}$: the first estimator is $\{\widehat{\mathbf{L}_i^{*}}, \widehat{\vect({\mathbf{Y}_{i}^{T}})}, \widehat{\Sigma}_{\mathbf{Y}_{i}}\}$ and the second estimator is $\{\widehat{\mathbf{L}_i^{*}}\mathbf{D}, \mathbf{D}^{-1}\widehat{\vect({\mathbf{Y}_{i}^{T}})}, \mathbf{D}^{-1}\widehat{\Sigma}_{\mathbf{Y}_{i}}(\mathbf{D}^{-1})^{T}\}$. 

To address this issue, we place a constraint on $\Sigma_{\mathbf{Y}}$ that requires its main diagonal element to be $1$.  In other words, the covariance matrix of latent factors $\Sigma_{\mathbf{Y}}$ is forced to be a correlation matrix by assuming the variance of factors to be $1$. This restriction has also been used in \cite{conti2014bayesian}, with multiple purposes: first, it ensures the uniqueness of $\Sigma_{\mathbf{Y}}$; in turn, it helps set the scale of $\vect({\mathbf{Y}_{i}^{T}})$ and consequently helps set the scale of $\mathbf{L}_i^{*}$.

\clearpage 

\section*{B. COVID-19 data application}
\subsection*{B.1 Reference grid approach}
The reference grid approach assumes the reference times as $\mathbf{t}_{\text{ref}}= \{0,7,14,...,49\}$, and replaces the actual observed time $t_{ij}$ (the $j$th time point of the $i$th subject) with the time from the reference grid, denoted as $t_{ij}^{\text{new}}$ below. Specifically, the replacement involves the following steps:
\begin{enumerate}
\item First, we fixed the first observed time point $t_{i1}$ as the value taken from the reference grid $t_{\text{ref}}$ (whichever is closet to $t_{i1}$); 
\item Second, for each of the remaining observed time points (i.e., $t_{ij}$, where $j = 2,...,q_i$): we calculated the corresponding time after transformation $t_{ij}^{\text{new}}$ using the formula $t_{ij}^{\text{new}}$ = $t_{i(j-1)}^{\text{new}} + (t_{ij} - t_{i(j-1)})$, aiming to keep the distance between adjacent times unchanged. If $t_{ij}^{\text{new}}$ belongs to the reference grid $t_{\text{ref}}$, keep it; if not, set $t_{ij}^{\text{new}}$ as the time in $t_{\text{ref}}$ that is closest to it. 
\end{enumerate}

\begin{figure}[hbt!]
\centering
\includegraphics[width=\textwidth]{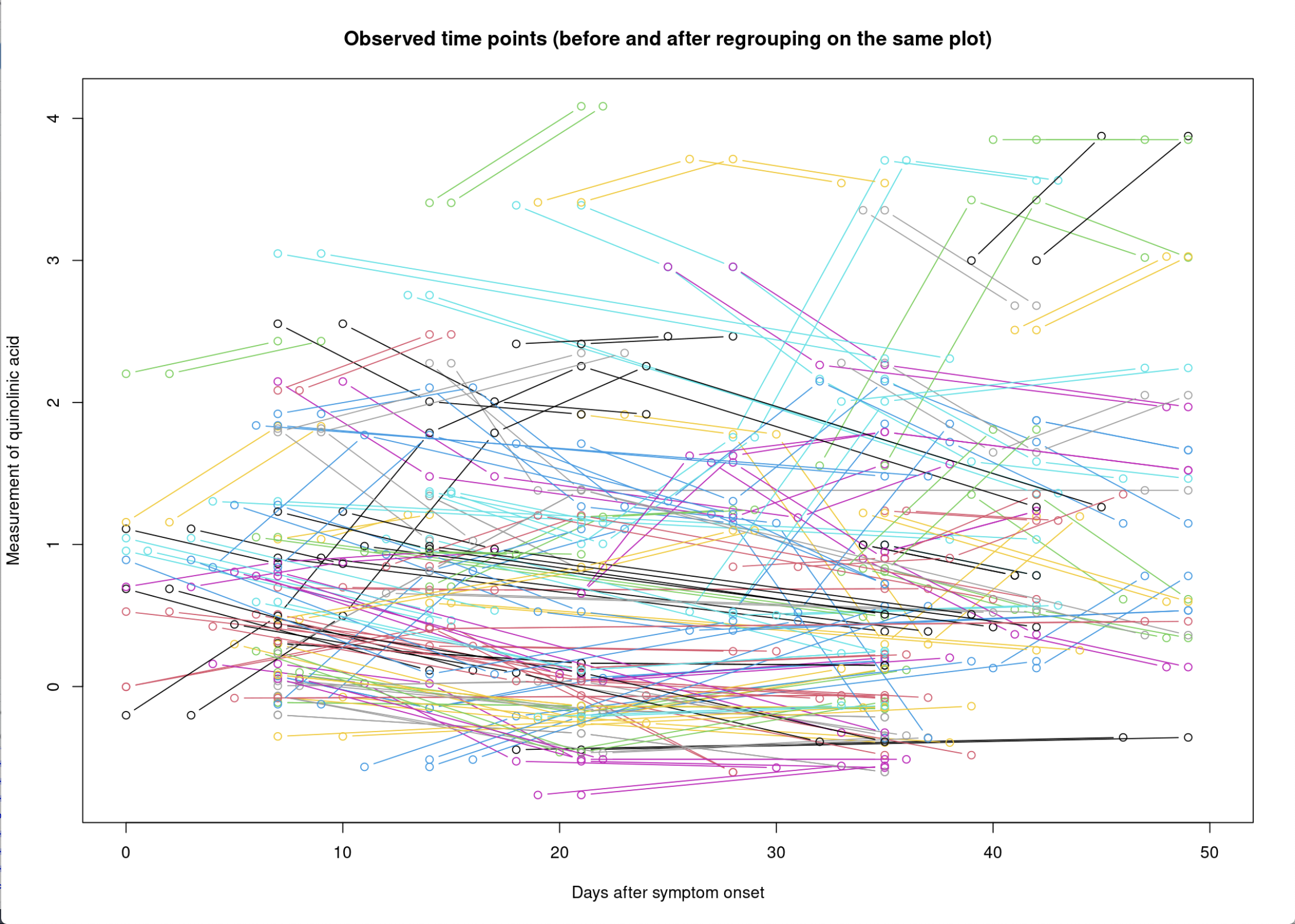}
\caption{Lines connect subject-specific observed times (before- and after- time transformation displayed alongside each other). Points denote the measurement of quinolinic acid at observation times. Note that all metabolites are measured at same times, we choose quinolinic acid as an example to draw the plots.}
\label{time_points_on_one_plots}
\end{figure}

Supplementary Figure~\ref{time_points_on_one_plots} shows subject-specific observed times before and after such transformation, taking the measurement of a specific biomarker (quinolic acid) as an example. The intra-class correlation coefficient (ICC) of distances between before- and after-transformation times is $0.908$, where class corresponds to individual. This number suggests a high correlation of distances within subjects, which reflects that before- and after-translation lines are roughly parallel to each other.

\clearpage

\begin{figure}[hbt!]
\centering
\includegraphics[width=\textwidth]{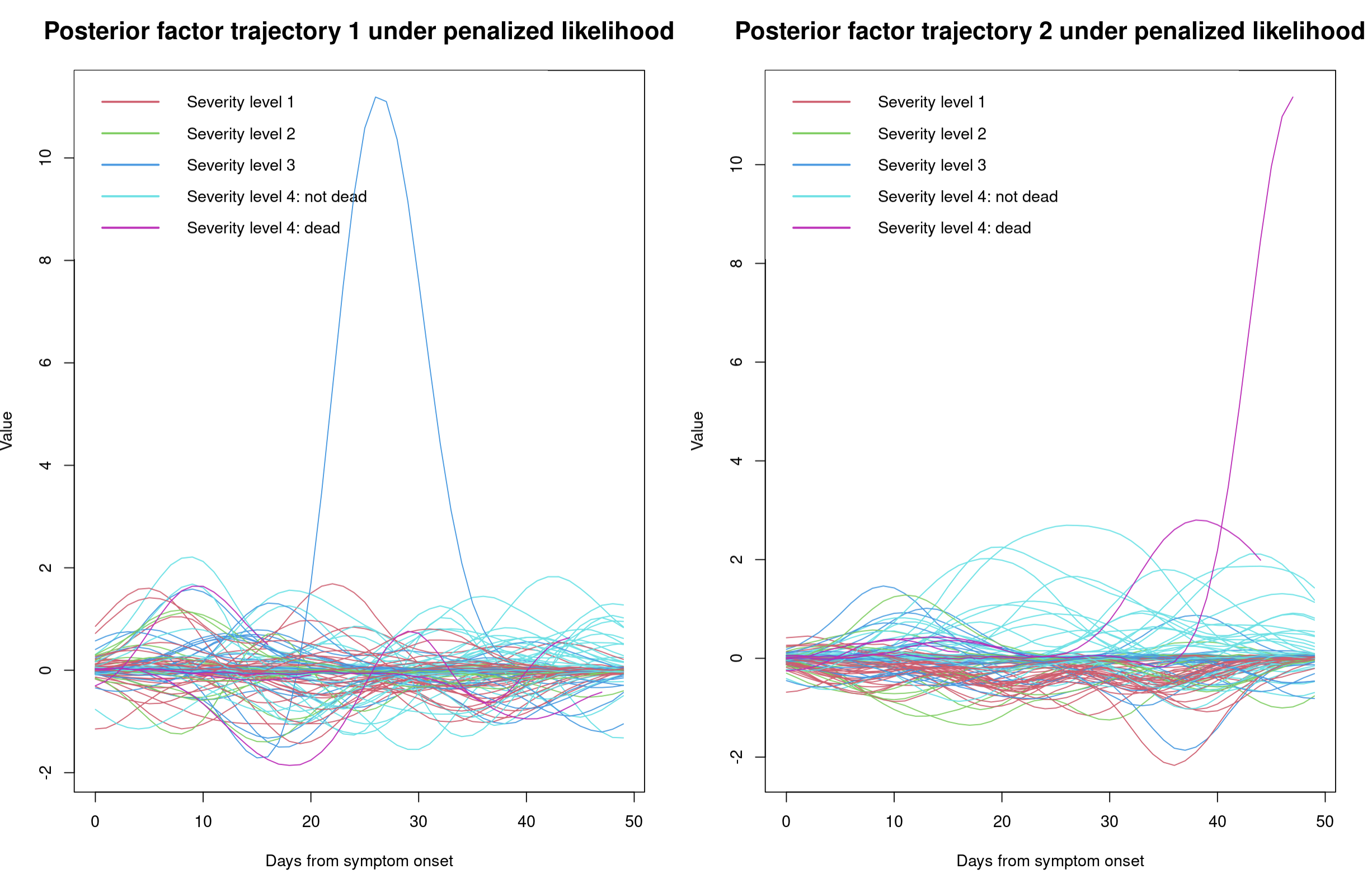}
\caption{Estimated subject-specific posterior factor trajectories for all patients (untruncated), using the exact method with $q = 50$. The number of latent factors is pre-specified as $k=2$. For people who did not survive for the full 7-week followup, we plotted their trajectories only before the death date; otherwise trajectories were plotted within a 7-week window after the onset of symptoms.}
\label{posterior_factor_trajectories_q_50}
\end{figure}

\begin{figure}[hbt!]
\centering
\includegraphics[width=\textwidth]{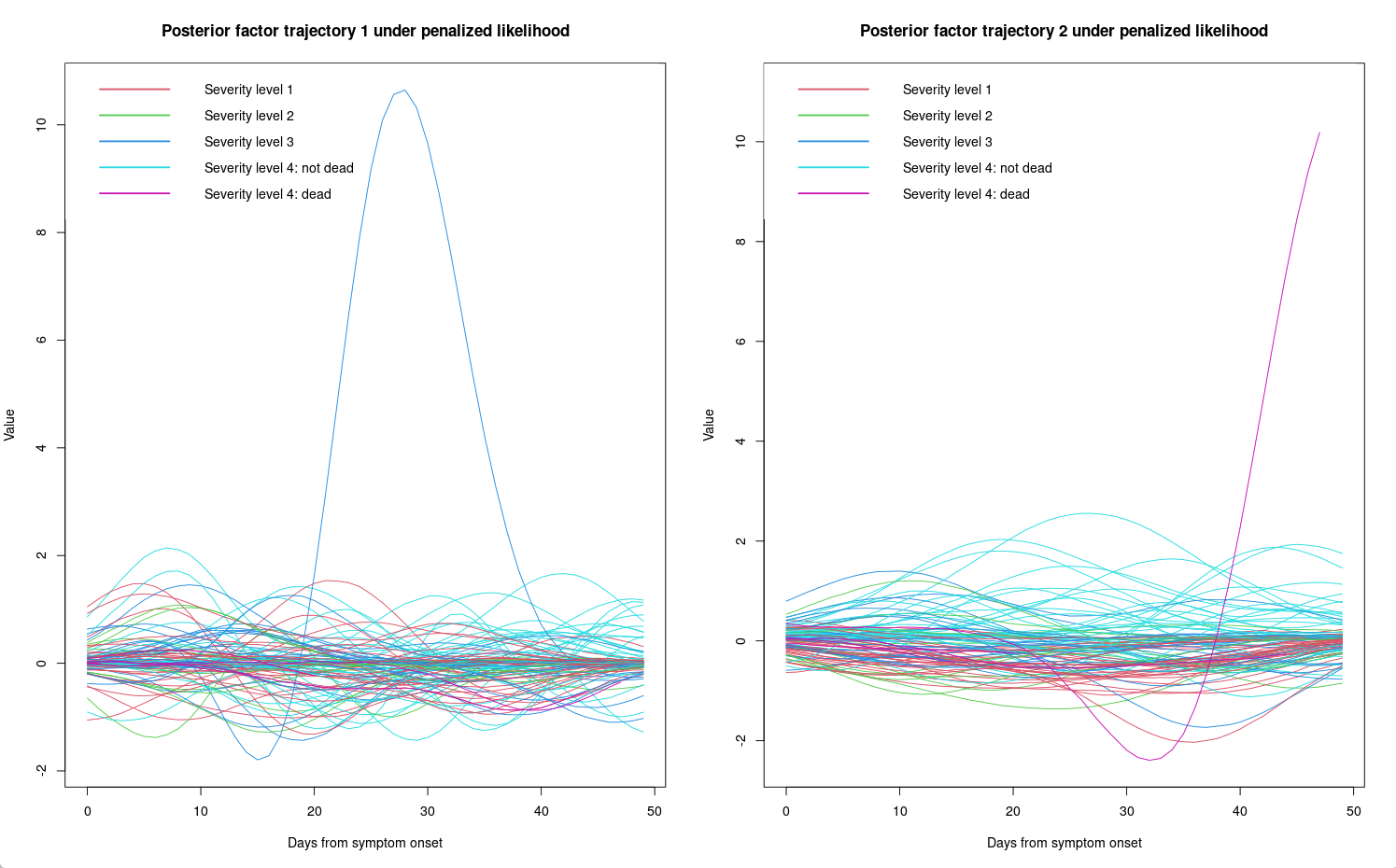}
\caption{Estimated subject-specific posterior factor trajectories for all patients (untruncated), using the approximation method with $q = 8$. The number of latent factors is pre-specified as $k=2$. For people who did not survive for the full 7-week followup, we plotted their trajectories only before the death date; otherwise trajectories were plotted within a 7-week window after the onset of symptoms.}
\label{posterior_factor_trajectories_q_8}
\end{figure}

We compare results using the exact method with $q = 50$ and the approximation method with $q = 8$ (both via our method). Taking the number of latent factors $k = 2$ as an example, Supplementary Tables~\ref{compare_loading_before_after_time_transformation_factor_1}-\ref{compare_loading_before_after_time_transformation_factor_2} show estimated factor loadings, and Supplementary Figure~\ref{posterior_factor_trajectories_q_50}-\ref{posterior_factor_trajectories_q_8} display estimated factor trajectories. Correspondence between metabolite indexes and metabolite names is displayed in Supplementary Table~\ref{code_metabolite}.


Both models led to similar results, which is as expected due to the reason described in Section 5.2 of the main manuscript (i.e., the specific design of the COVID-19 study). Note that the individual corresponding to the outlier trajectory with extremely high values (around $10$ on day $48$) has the largest BMI ($37.46$) among all patients in the study, which may explain the unusual pathway activity detected by our approach.

\clearpage 

\setlength{\LTcapwidth}{\linewidth}
\begin{longtable}{c|c|c}

\caption{Comparison of estimated loadings on factor 1, between the exact method with $q = 50$ and the approximation method with $q = 8$. The displayed estimate is the median of the posterior samples. Supplementary Table~\ref{code_metabolite} displays metabolite names corresponding to metabolite indexes.} 

\label{compare_loading_before_after_time_transformation_factor_1} \\

\hline
Metabolite index & Exact method with $q = 50$ & Approximation method with $q = 8$\\

\hline
 1& -0.03& -0.04 \\
2 & 0.00 & 0.00\\
3& 0.00 & 0.00\\
4& 0.00 & 0.00 \\
5& 0.09 & 0.10 \\
6& 0.00 & 0.00\\
7& 0.00 & 0.00 \\
8& 0.00 & 0.00 \\
9& 0.00 & 0.00 \\
10& 0.00 & 0.00 \\
11& -0.11 & -0.12 \\ 
12& -0.11 & -0.12 \\ 
13& -0.15 & -0.17 \\ 
14& -0.48 & -0.53 \\
15& -0.33 & -0.37 \\ 
16& -0.39 & -0.43 \\ 
17& -0.41 & -0.45 \\
18& -0.33 & -0.36 \\ 
19& -0.29 & -0.32 \\
20& -0.29 & -0.32 \\ 
21& -0.31 & -0.34 \\ 
22& -0.32 & -0.35 \\
23& -0.33 & -0.36 \\ 
24& -0.28 & -0.31 \\ 
25& -0.49 & -0.54 \\  
26& -0.25 & -0.27 \\ 
27& -0.41  & -0.45 \\ 
28& -0.39 & -0.42 \\ 
29& -0.24 & -0.26 \\
30& -0.42 & -0.46 \\ 
31& -0.38 & -0.41 \\ 
32& -0.22 & -0.24 \\
33& -0.39 & -0.42 \\
34& -0.39 & -0.43 \\ 
35& -0.31 & -0.34 \\ 

\hline 
\end{longtable}

\clearpage

\setlength{\LTcapwidth}{\linewidth}
\begin{longtable}{c|c|c}

\caption{Comparison of estimated loadings on factor 2, between the exact method with $q = 50$ and the approximation method with $q = 8$. The displayed estimate is the median of the posterior samples. Supplementary Table~\ref{code_metabolite} displays metabolite names corresponding to metabolite indexes.} 

\label{compare_loading_before_after_time_transformation_factor_2} \\

\hline
Metabolite index & Exact method with $q = 50$ & Approximation method with $q = 8$\\

\hline
 1& 0.63 & 0.67 \\
2 & 0.80 & 0.86 \\
3& 0.50 & 0.54 \\
4& 0.00 & 0.00 \\
5& 0.89 & 0.95 \\
6& 0.64 & 0.69 \\
7& 0.79 & 0.85 \\
8& 0.41 & 0.44 \\
9& 1.22 & 1.29 \\
10& 0.00 & 0.00 \\
11& -0.17 & -0.18 \\ 
12& 0.43 & 0.45  \\ 
13& 0.24 & 0.26 \\ 
14& -0.08 & -0.09 \\
15& 0.00 & 0.00 \\ 
16& 0.00 & 0.00 \\ 
17& -0.03 & -0.02 \\
18& 0.00 & 0.00 \\ 
19& 0.00 & 0.00 \\
20& 0.00 & 0.00 \\ 
21& 0.00 & 0.00 \\ 
22& -0.06 & -0.06 \\
23& 0.08 & 0.09 \\ 
24& 0.05 & 0.06 \\ 
25& 0.00 & 0.00 \\  
26& 0.07 & 0.08 \\ 
27& 0.00  & 0.00 \\ 
28& -0.12 & -0.13 \\ 
29& -0.25 & -0.28 \\
30& 0.00 & 0.00 \\ 
31& 0.00 & 0.00 \\ 
32& 0.00 & 0.00 \\
33& 0.04 & 0.05 \\
34& 0.00 & 0.00 \\ 
35& 0.00 & 0.00 \\
\hline 

\end{longtable}

\clearpage 

\setlength{\LTcapwidth}{\linewidth}
\begin{longtable}{c|c}

\caption{Correspondence between metabolite index and metabolite name.} 

\label{code_metabolite} \\

\hline
Metabolite index & Metabolite name \\
\hline
1 & 3-hydroxyanthranilic acid \\ 
2 & 3-hydroxykynurenine \\ 
3 & 5-hydroxyindoleacetic acid \\ 
4 & indole-3-acetic acid \\ 
5 & kynurenic acid \\ 
6 & kynurenine \\ 
7 & neopterin \\ 
8 & picolinic acid \\ 
9 & quinolinic acid \\ 
10 & serotonin \\ 
11 & tryptophan \\ 
12 & xanthurenic acid \\ 
13 & 1-methylhistidine \\ 
15 & alpha-aminobutyric acid \\ 
16 & Arginine \\ 
17 & Asparagine \\ 
18 & Citrulline \\ 
19 & Glutamic acid \\ 
20 & Glutamine \\ 
21 & Glycine \\ 
22 & Histidine \\ 
23 & Isoleucine \\ 
24 & Leucine \\ 
25 & Methionine \\ 
26 & Phenylalanine \\ 
27 & Proline \\
28 & Serine \\
29 & Taurine \\ 
30 & Threonine \\ 
31 & Tyrosine \\ 
32 & Aspartic acid \\ 
33 & Lysine \\ 
34 & Ornithine \\ 
35 & Valine \\ 

\hline 
\end{longtable}

\clearpage 

\clearpage 

\subsection*{B.2 Results under the number of latent factors $k = 2$}

\begin{figure}[htp]
\centering
\includegraphics[width = 0.8\textwidth]{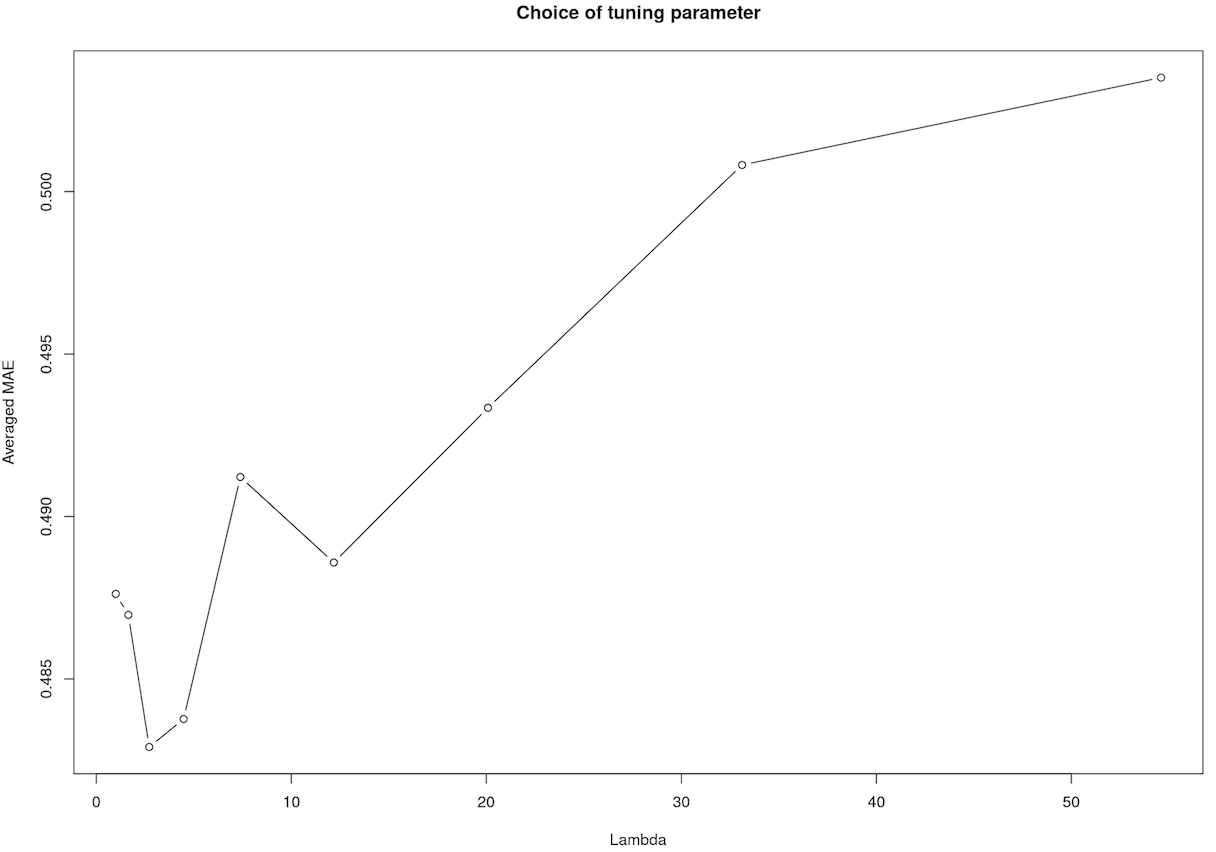}
\caption{Cross validation results for choosing the tuning parameter $\lambda$. We vary $\text{ln}(\lambda)$ from $-4$ to $4$, with the length of step as $0.5$; this results in $17$ candidate values of $\lambda$, from $0.02$ to $54.60$. MAE stands for mean absolute error between estimated biomarker expressions and the truth.}
\label{choice_of_tuning_parameter}
\end{figure}

\begin{figure}[htp]
\centering
\includegraphics[width=\textwidth]{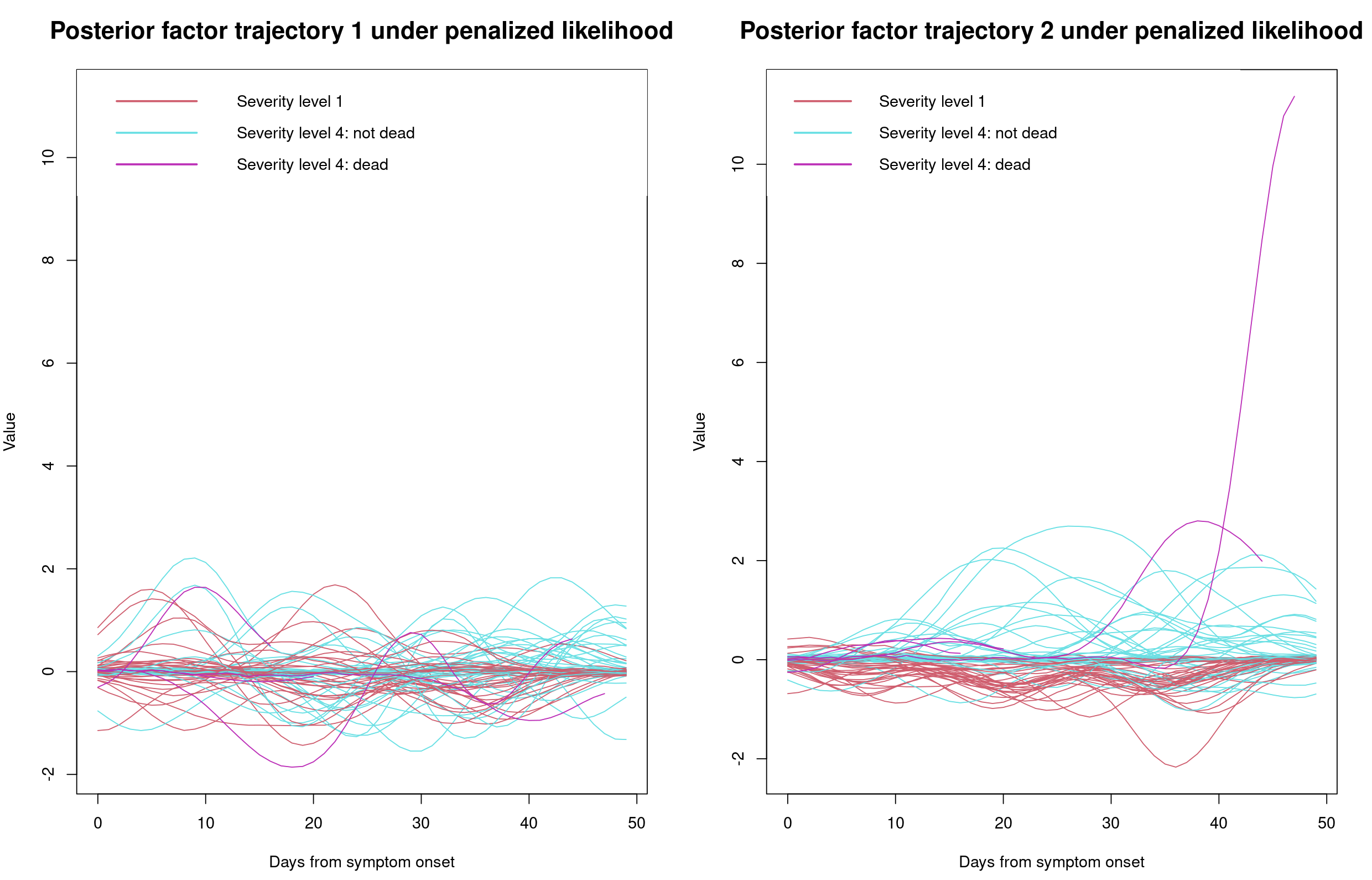}
\caption{Estimated subject-specific posterior factor trajectories for patients with severity levels 1 and 4 only (untruncated), using the exact method with $q = 50$. The number of latent factors is pre-specified as $k=2$. For people who did not survive for the full 7-week followup, we plotted their trajectories only before the death date; otherwise trajectories were plotted within a 7-week window after the onset of symptoms.}
\end{figure}

\clearpage 

\subsection*{B.3 Results under the number of latent factors $k = 3$}
\begin{figure}[htp]
\centering
\includegraphics[width=\textwidth]{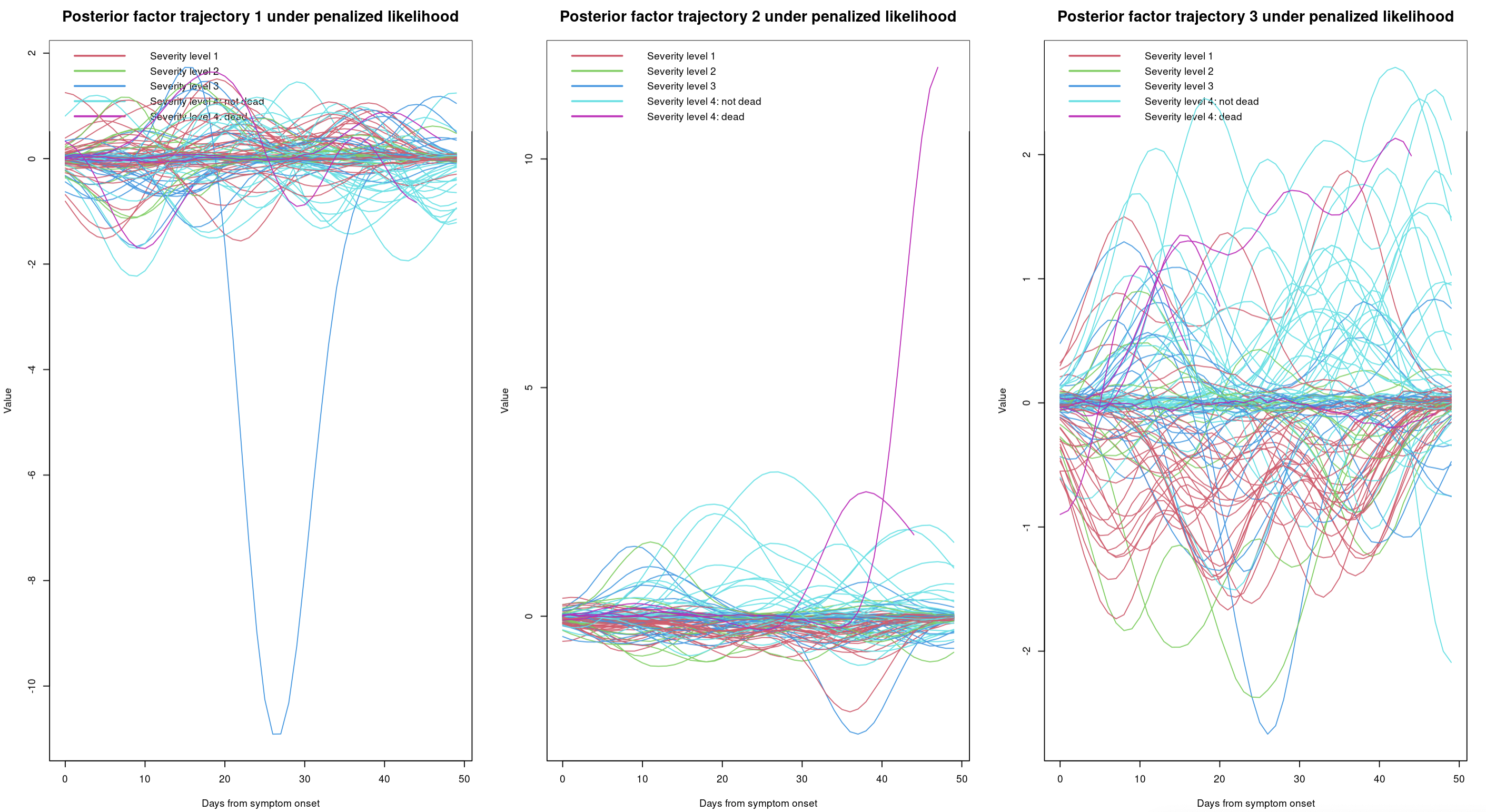}
\caption{Estimated subject-specific posterior factor trajectories for all patients (untruncated), using the exact method with $q = 50$. The number of latent factors is pre-specified as $k=3$. For people who did not survive for the full 7-week followup, we plotted their trajectories only before the death date; otherwise trajectories were plotted within a 7-week window after the onset of symptoms.}
\end{figure}

\begin{figure}[htp]
\centering
\includegraphics[width=\textwidth]{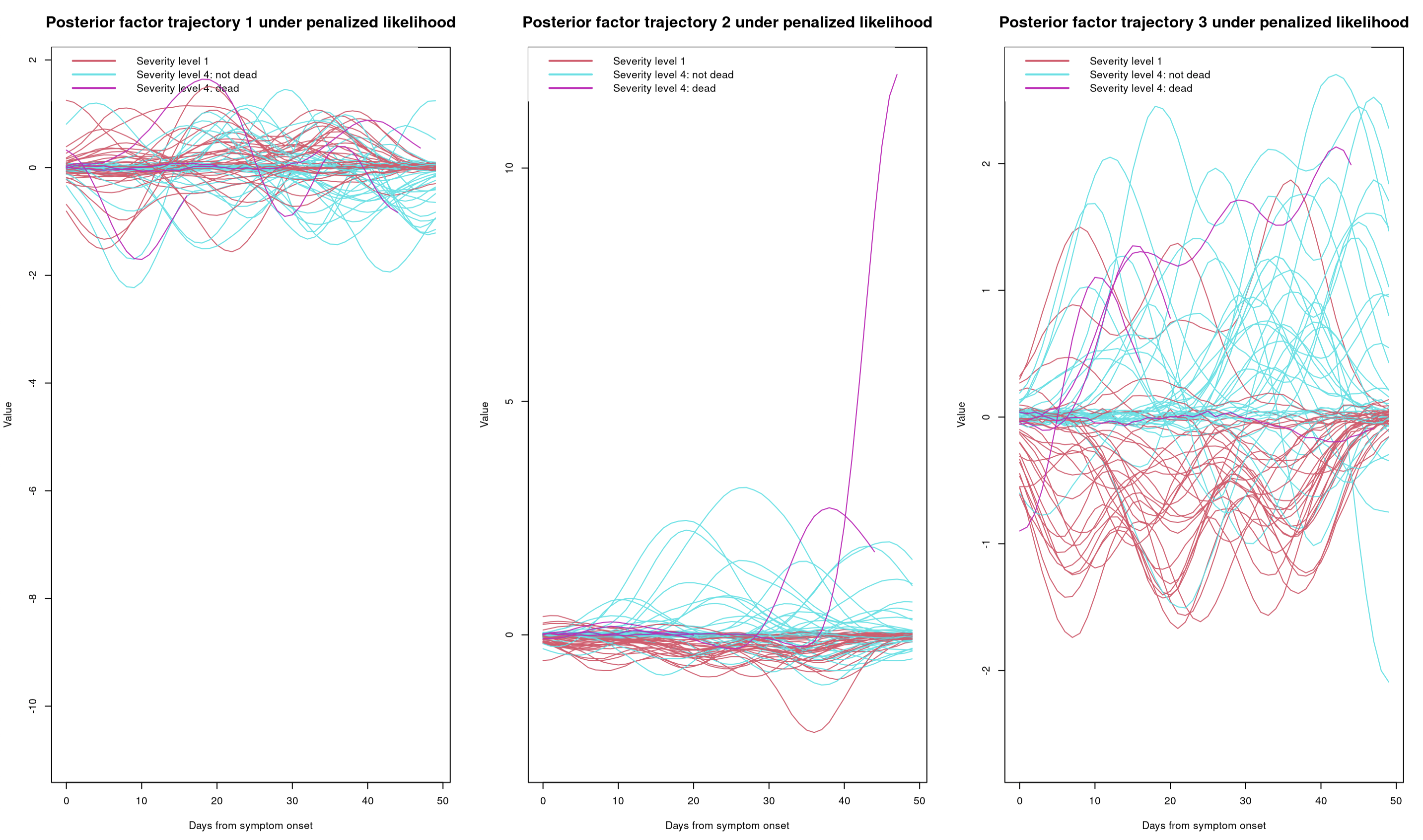}
\caption{Estimated subject-specific posterior factor trajectories for patients with severity levels 1 and 4 only (untruncated), using the exact method with $q = 50$. The number of latent factors is pre-specified as $k=3$. For people who did not survive for the full 7-week followup, we plotted their trajectories only before the death date; otherwise trajectories were plotted within a 7-week window after the onset of symptoms.}
\end{figure}

\clearpage 

\subsection*{B.4. Standardization of input}
When fitting the model to the real-data, we standardized the input of MOGP (i.e., time in our case) so that the range of inputs after standardization matches that in the R package GPFDA. Specifically, we divided the original times (range between 0 and 49) by 49, resulting in the after-transformation times ranging between 0 and 1. When displaying the results, we use the original time for x-axis to facilitate interpretation (i.e., with `day' as the unit).

\section*{C. Simulation}
\begin{figure}[htp]
\centering
\includegraphics[width=0.5\textwidth]{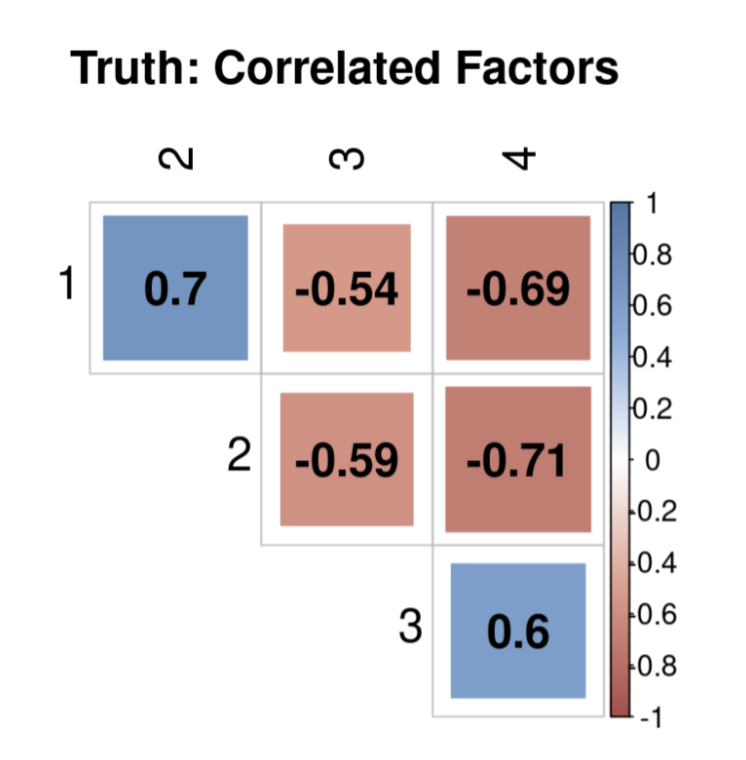}
\caption{True cross-correlations among the $4$ latent factors in the simulation study.}
\end{figure}

\end{document}